\documentclass[longauth]{aa} 
\usepackage{lscape}
\usepackage{natbib,twoopt}
\usepackage{graphicx}
\usepackage[usenames,dvipsnames]{xcolor}
\usepackage{multirow}
\usepackage[flushleft]{threeparttable}
\usepackage{bm}
\usepackage{txfonts}

\begin{document} 

   \title{The eROSITA Final Equatorial-Depth Survey (eFEDS)}
   \subtitle{X-ray stacking analysis of Subaru's optically selected clusters spanning low richness regime}
    
   \author{N. T. Nguyen-Dang \inst{1,2}
          \and
          N. Ota\inst{3, 2}\fnmsep\thanks{Corresponding author: naomi@astro.uni-bonn.de}
          \and
          N. Okabe\inst{4,5,6}
          \and
          M. Oguri\inst{7,8,9}
          \and 
          I. Mitsuishi\inst{10}
          \and
          T. H. Reiprich\inst{3}
          \and
          F. Pacaud\inst{3}
          \and
          E. Bulbul\inst{11}
          \and
          J. S. Sanders\inst{11}
          \and
          M. Br\"uggen\inst{12}
          \and
          A. Liu\inst{13}
          \and
          Y. Tsujita\inst{2}
          \and
          I. Chiu\inst{14,15,16}
          \and 
          V. Ghirardini\inst{11, 17}
          \and
          S. Grandis\inst{18}
          \and
          M. Klein\inst{19}
          \and
          K. Migkas\inst{20,21}
          \and
          H. Miyatake\inst{9,22}
          \and
          S. Miyazaki\inst{23,24}
          \and
          M. E. Ramos-Ceja\inst{11}
          }

        \institute{ Institut f\"{u}r Astronomie und Astrophysik T\"{u}bingen (IAAT), Universit\"{a}t T\"{u}bingen, Sand 1, 72076 T\"{u}bingen, Germany\\
        \email{nguyen@astro.uni-tuebingen.de}
        \and
            Department of Physics, Nara Women's University, Kitauoyanishi-machi, Nara, 630-8506, Japan
        \and
            Argelander-Institut f\"{u}r Astronomie (AIfA), Universit\"{a}t Bonn, Auf dem H\"{u}gel 71, 53121 Bonn, Germany
        \and
            Physics Program, Graduate School of Advanced Science and Engineering, Hiroshima University, 1-3-1 Kagamiyama, Higashi-Hiroshima, Hiroshima 739-8526, Japan
        \and
            Hiroshima Astrophysical Science Center, Hiroshima University, 1-3-1 Kagamiyama, Higashi-Hiroshima, Hiroshima 739-8526, Japan
        \and
            Core Research for Energetic Universe, Hiroshima University, 1-3-1, Kagamiyama, Higashi-Hiroshima, Hiroshima 739-8526, Japan 
        \and 
            Center for Frontier Science, Chiba University, 1-33 Yayoi-cho, Inage-ku, Chiba 263-8522, Japan
        \and 
            Department of Physics, Graduate School of Science, Chiba University, 1-33 Yayoi-Cho, Inage-Ku, Chiba 263-8522, Japan
        \and 
            Kavli Institute for the Physics and Mathematics of the Universe (Kavli IPMU, WPI), University of Tokyo, Chiba 277-8582, Japan
        \and
            Graduate School of Science, Division of Particle and Astrophysical Science, Nagoya University, Furo-cho, Chikusa-ku, Nagoya, Aichi, 464-8602, Japan
        \and         
            Max Planck Institute for Extraterrestrial Physics, Giessenbachstrasse 1, 85748 Garching, Germany
        \and
            University of Hamburg, Hamburger Sternwarte, Gojenbergsweg 112, 21029 Hamburg, Germany
        \and
            Institute for Frontiers in Astronomy and Astrophysics, Beijing Normal University, Beijing 102206, China
        \and
            Tsung-Dao Lee Institute, and Key Laboratory for Particle
            Physics, Astrophysics and Cosmology, Ministry of Education,
            Shanghai Jiao Tong University, Shanghai 200240, China
        \and
            Department of Astronomy, School of Physics and Astronomy,
            and Shanghai Key Laboratory for Particle Physics and Cosmology,
            Shanghai Jiao Tong University, Shanghai 200240, China
        \and
            Academia Sinica Institute of Astronomy and Astrophysics (ASIAA), 11F of AS/NTU Astronomy-Mathematics Building, No.1, Sec. 4, Roosevelt Rd, Taipei10617, Taiwan
        \and
            INAF, Osservatorio di Astrofisica e Scienza dello Spazio, via Piero Gobetti 93/3, I-40129 Bologna, Italy
        \and
            Universit\"{a}t Innsbruck,  Institut f\"{u}r Astro- und Teilchenphysik, Technikerstr. 25/8, 6020 Innsbruck, Austria
        \and
            Faculty of Physics, Ludwig-Maximilians-Universit{\"a}t, Scheinerstr. 1, 81679, Munich, Germany
        \and
            Leiden Observatory, Leiden University, PO Box 9513, 2300 RA Leiden, the Netherlands
        \and
            SRON Netherlands Institute for Space Research, Niels Bohrweg 4, NL-2333 CA Leiden, the Netherlands 
        \and
            Kobayashi-Maskawa Institute for the Origin of Particles and the Universe (KMI), Nagoya University, Nagoya, 464-8602, Japan
        \and
            National Astronomical Observatory of Japan, 2-21-1 Osawa, Mitaka, Tokyo 181-8588, Japan
        \and
            SOKENDAI (The Graduate University for Advanced Studies), Mitaka, Tokyo, 181-8588, Japan
            }

   \date{Received September 15, 1996; accepted March 16, 1997}
 
  \abstract
{\textit{Context.} This is the second paper in a series exploring the X-ray properties of galaxy clusters optically selected by the Subaru Hyper Suprime-Cam (HSC) survey, using data from the SRG/eROSITA Final Equatorial-Depth Survey (eFEDS).

\textit{Aims.} We aim to investigate scaling relations between observable cluster properties and mass, and to study the radial X-ray profiles of a large sample of optically selected clusters.

\textit{Methods.} We analyze a sample of 997 CAMIRA clusters with richness $N > 15$ and redshifts of $0.1 < z < 1.3$. Using bolometric luminosities derived from count rates and a weak-lensing mass calibration, we study the $L-M$ and $N-M$ scaling relations through stacking analysis, while accounting for selection effects and redshift evolution. We also compare clusters with and without X-ray counterparts in the eFEDS catalog in terms of their scaling relations and surface brightness profiles.

\textit{Results.} The best-fit $L-M$ slope ($1.56^{+0.14}_{-0.12}$) is slightly steeper than the self-similar prediction, yet remains consistent with our previous findings. The $N-M$ slope ($0.766^{+0.070}_{-0.060}$) broadly agrees with theoretical expectations and other optical samples. The data do not require any additional redshift evolution beyond the standard self-similar scaling, although current constraints on evolution remain weak. X-ray detected clusters exhibit a marginally steeper $L-M$ slope, higher central surface brightness, and more centrally concentrated X-ray profiles than undetected systems.

\textit{Conclusions.} Our results highlight systematic differences in the X-ray properties between optically and X-ray selected cluster samples. This study extends scaling relation analyses into lower mass and luminosity regimes, demonstrating the value of combining deep X-ray and optical surveys like eROSITA and Subaru HSC.

}


   \maketitle

\section{Introduction}
Galaxy clusters are fundamental tracers of cosmic structure formation and evolution, and they provide key constraints on cosmology (see e.g., \citealt{Pratt19} for a review). Building comprehensive cluster samples requires multi-wavelength selection methods, ranging from sub-millimeter to X-ray data, each probing different physical components of clusters. Clusters can be identified through optical, weak-lensing, and X-ray surveys, each subject to distinct selection effects. Optical methods trace overdensities of red-sequence galaxies \citep{2014ApJ...783...80R, 2014MNRAS.444..147O, 2019MNRAS.485..498M, 2023A&A...675A.202V}, weak lensing (WL) probes the gravitational potential \citep{2018PASJ...70S..27M, 2021PASJ...73..817O, 2025OJAp....8E...2C, 2024OJAp....7E..90C}, and X-ray surveys identify the hot intracluster medium (ICM) via thermal bremsstrahlung emission \citep{2001A&A...369..826B, 2017MNRAS.469.3738S, 2022A&A...658A..59X}.

Because these methods probe different cluster properties, their resulting samples exhibit systematic differences \citep{2020PASJ...72....1O, 2023A&A...669A.110O, 2021MNRAS.503.5624W}. X-ray flux-limited samples tend to be biased toward massive, relaxed clusters with cool cores \citep{2007MNRAS.382.1289P, 2010A&A...513A..37H, 2011A&A...536A..37A, 2011A&A...526A..79E, 2018A&A...619A.162X}. In contrast, optically selected clusters generally include lower-mass and more dynamically disturbed systems with more irregular X-ray morphologies. They may also be affected by projection effects, which can bias richness estimates and increase scatter in observable--mass relations \citep{2019MNRAS.482..490C}. Several studies have reported that optical samples contain a higher fraction of disturbed systems and exhibit shallower or more scattered scaling relations than X-ray selected clusters \citep[e.g.,][]{2023A&A...669A.110O}.

These systematic differences highlight the need to understand and properly account for selection effects when interpreting scaling relations between observables and mass. Self-similar models \citep{Kaiser86} provide a theoretical baseline for the expected form of these relations, although non-gravitational processes such as AGN feedback, radiative cooling, and mergers can drive deviations from the self-similar predictions \citep{2009A&A...498..361P, 2020MNRAS.494..161M,2013SSRv..177..247G}. Scaling relations examined across a broad redshift range also test baryonic evolution, yet published results on redshift evolution remain mixed \citep{2007ApJ...668..772M, 2007MNRAS.382.1289P, 2009ApJ...692.1033V, 2018MNRAS.478.3072C, 2022A&A...661A...7B,2011A&A...535A...4R}.

The eROSITA X-ray telescope aboard the SRG mission \citep{Predehl21, 2021A&A...656A.132S} is conducting an all-sky survey, providing unprecedented coverage of low-mass and intermediate-redshift clusters. The eROSITA Final Equatorial-Depth Survey (eFEDS; \citealt{2022A&A...661A...1B}) overlaps with deep optical surveys, including SDSS, Legacy, and especially Subaru Hyper Suprime-Cam (HSC). The latter delivers high-quality optical cluster catalogs \citep{2018PASJ...70S..20O} and WL mass measurements \citep{2018PASJ...70S..27M}, enabling joint optical--X-ray studies of large cluster samples \citep{Liu22, Bulbul22}.

Our previous work \citep{2023A&A...669A.110O} analyzed the X-ray and morphological properties of a small set of 43 high-richness ($N > 40$) CAMIRA clusters in the eFEDS field. In the present study, we extend this analysis to a much larger and lower-richness sample of 997 clusters with $N > 15$ and $0.1 < z < 1.3$. This substantially increases the statistical power to probe the X-ray properties of optically selected clusters and for exploring scaling relations down to lower masses and luminosities.

This work addresses two key questions: (1) How do the average X-ray properties differ between optical clusters with and without X-ray counterparts in the eFEDS catalog? (2) What are the observable--mass scaling relations of optically selected clusters over a broad mass and redshift range, and how do they compare with self-similar expectations?

The structure of this paper is as follows. Sect.~\ref{sec:sample} describes the construction of the optical cluster sample. Sect.~\ref{sec:xray} presents the X-ray analysis, including stacked luminosities and radial profiles. The WL mass calibration is summarized in Sect.~\ref{sec:weaklens}. Sect.~\ref{sec:results} reports the scaling relations and surface-brightness results, followed by discussion in Sect.~\ref{sec:discussion}. Sect.~\ref{sec:summary} provides a summary of our findings.

We adopt a flat $\Lambda$CDM cosmology with $\Omega_\mathrm{M} = 0.3$, $\Omega_\Lambda = 0.7$, and $h = 0.7$. X-ray spectral modeling uses the abundance table from \cite{2009ARA&A..47..481A}. Errors are quoted at the 68\% confidence level unless otherwise noted.

\section{Sample}\label{sec:sample}

\begin{figure}
     \centering
     \includegraphics[width=0.48\textwidth]{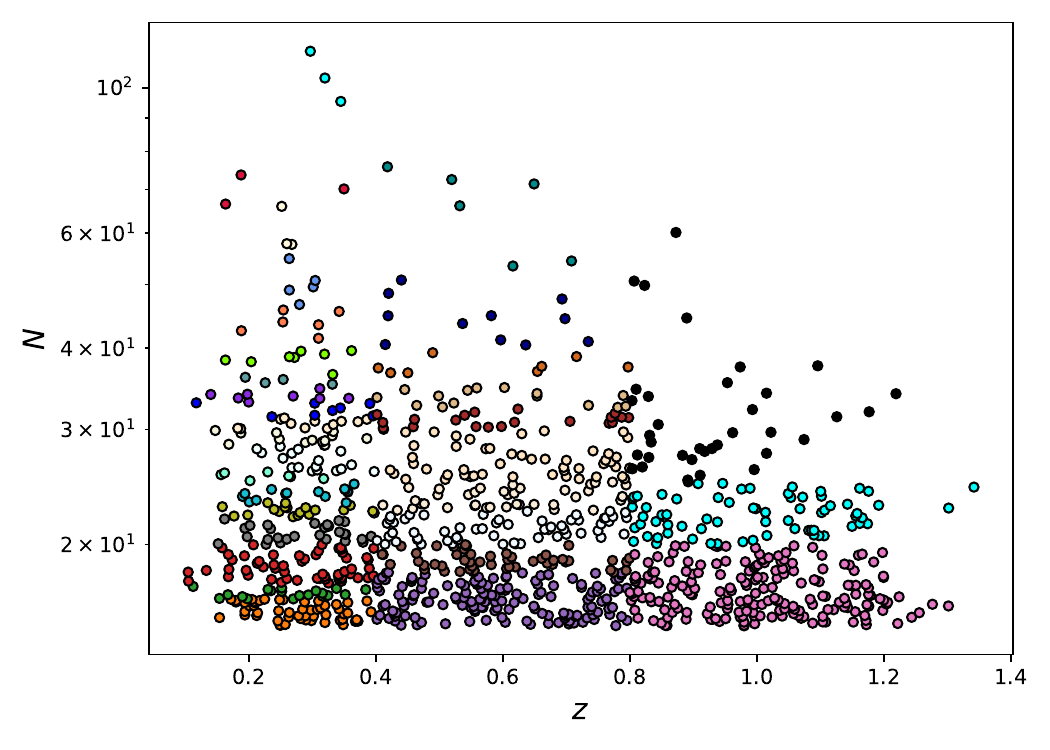}
    \caption{Distribution of the CAMIRA clusters in the richness--redshift plane. Each point represents a cluster. Colors indicate the stacked groups used in the X-ray stacking analysis, defined by similar richness and redshift ranges (see Sects.~\ref{sec:sample} and~\ref{subsec:stack_lumi}).}\label{fig:stacking}
 \end{figure}

Out of the 21,250 optically selected galaxy clusters in the CAMIRA S20A v2 catalog \citep{2018PASJ...70S..20O}, we selected 997 systems located within the eFEDS field. The sample spans a richness range of $15 < N < 114$ and a redshift range of $0.10 < z < 1.34$ (Fig.~\ref{fig:stacking}). Table~\ref{tab:sample} lists the first ten clusters; the full table is available at the CDS.

While \citet{2023A&A...669A.110O} analyzed only the 43 high-richness clusters ($N > 40$), our study includes all 997 objects. This greatly increases the statistical power for X-ray follow-up of optically selected clusters and enables a systematic investigation across a broader mass range.

Higher-richness clusters are more likely to be detected in X-rays. In our sample, only 171 clusters (17\%) are detected and listed in the eFEDS X-ray catalog \citep{Liu22}. Following our earlier work \citep{2023A&A...669A.110O}, we identified X-ray counterparts by searching for spatially extended X-ray sources within the characteristic radius $R_{500}$\footnote{The radius within which the mean density is 500 times the critical density of the Universe at the cluster redshift.} of each CAMIRA center, requiring a redshift consistency of $|\Delta z| < 0.02$. This threshold is comparable to the typical photometric-redshift uncertainty for cluster surveys such as SDSS \citep[e.g.,][]{2013A&A...558A..75T}. Although it is more conservative than the CAMIRA photometric-redshift precision of $\Delta z / (1+z) < 0.01$ \citep{2018PASJ...70S..20O}, it ensures robust association between the optical and X-ray detections.

\begin{table*}[hbt]
    \caption{Sample list.}\label{tab:sample}
    \centering
    \begin{tabular}{llllllllllllll}  \hline \hline
    
    Cluster & $z$ & $\hat{N}_{\mathrm{mem}}$\tablefootmark{(a)} & $R_{500}$ (Mpc/$''$) & RA, Dec (deg) & $L$\tablefootmark{(b)} & $L_{\mathrm{err}}$\tablefootmark{(c)} \\
    \hline

HSC J083151+031637 & 1.224 & 16.6 & 0.403 / 48 & 127.9634 , 3.2769 & 0.0 & 46.7 \\
HSC J083303+024433 & 0.187 & 30.1 & 0.725 / 231 & 128.2616 , 2.7425 & 17.6 & 3.3 \\
HSC J083317-003730 & 0.417 & 18.1 & 0.534 / 96 & 128.3223 , -0.6250 & 0.0 & 5.8 \\
HSC J083321+033847 & 0.773 & 24.9 & 0.559 / 75 & 128.3369 , 3.6465 & 10.5 & 21.0 \\
HSC J083322+031634 & 0.826 & 22.8 & 0.527 / 69 & 128.3410 , 3.2761 & 0.0 & 20.7 \\
HSC J083322-011142 & 0.289 & 28.7 & 0.689 / 158 & 128.3414 , -1.1949 & 44.2 & 6.6 \\
HSC J083326+002545 & 0.794 & 18.2 & 0.479 / 63 & 128.3593 , 0.4292 & 26.4 & 21.0 \\
HSC J083338+033453 & 0.951 & 18.9 & 0.464 / 58 & 128.4085 , 3.5813 & 176.7 & 47.6 \\
HSC J083339-011229 & 1.068 & 22.7 & 0.489 / 59 & 128.4140 , -1.2081 & 20.5 & 42.0 \\
HSC J083351+004427 & 0.380 & 20.2 & 0.568 / 109 & 128.4621 , 0.7408 & 7.8 & 6.2 \\
\hline
    \end{tabular}
    \tablefoot{
    The first ten CAMIRA clusters are listed here for illustration, while the complete list is available online.
\tablefoottext{a}{Richness.} \tablefoottext{b}{Bolometric luminosity within the scale radius $R_{500}$, in units of $10^{42}~{\rm erg\,s^{-1}}$}. \tablefoottext{c} $1\sigma$ error or upper limit of luminosity.}
\end{table*}

\section{X-ray data analysis} \label{sec:xray}

\subsection{Data reduction}  
\label{subsec:data_reduction}

We processed the eFEDS data from the seven telescope modules (TMs) using version \texttt{eSASSusers\_201009} of the eROSITA Science Analysis Software System (eSASS) \citep{2022A&A...661A...1B}, following the procedure described in \citet{2023A&A...669A.110O}. Cleaned event files were extracted, and point sources were removed using the main eFEDS X-ray source catalog\footnote{\url{https://erosita.mpe.mpg.de/edr/eROSITAObservations/Catalogues/}} \citep{2022A&A...661A...1B}.

\subsection{Stacking analysis of X-ray luminosity}\label{subsec:stack_lumi}

Cluster source and background regions were defined by a circular region of radius $R_{500}$ centered on each CAMIRA cluster, and an annular background region with inner and outer radii of 2.5 and 4.0~Mpc, respectively. Following \citet{2023A&A...669A.110O}, $R_{500}$ is computed from the mass--richness relation of \citet{2019PASJ...71...79O} using the richness values in the CAMIRA catalog \citep{2018PASJ...70S..20O}.

Because most clusters have low photon statistics, we estimate the bolometric (0.01--30.0~keV) X-ray luminosity from count rates rather than performing individual spectral fits. The bolometric luminosity is given by
\[
L_{\rm bol} = \left( \frac{L_{\rm model}}{C_{\rm model}} \right) C_{\rm obs},
\]
where $C_{\rm obs}$ is the observed 0.5--2.0~keV count rate and $(L_{\rm model}/C_{\rm model})$ is a conversion factor computed from an APEC thin-thermal plasma model \citep{Smith01, Foster12} with a metallicity of 0.3~solar, corrected for Galactic absorption using the \texttt{tbabs} model \citep{Wilms00}. Instrumental response files were generated at each cluster position using the eROSITA \texttt{srctool}. Gas temperatures were estimated from the richness--temperature relation for optically selected clusters \citep{2018PASJ...70S..20O}.

The 0.5--2.0~keV count rate is measured from the merged event file of all seven TMs, including TM5 and TM7, which may be affected by optical light leakage under certain observing conditions \citep{Predehl21}. To assess the robustness of our results, we also computed luminosities using only the five unaffected modules (TM1, TM2, TM3, TM4, TM6). The close agreement between the two measurements demonstrates that optical leakage has a negligible impact on our analysis.

To increase statistical significance, we perform a stacking analysis. Clusters are grouped by similar richness and redshift, ensuring that each group contains at least two clusters and a total of approximately 1000 X-ray photon counts. The sample is first divided into redshift bins ($z < 0.4$, $0.4 \leq z < 0.8$, and $z \geq 0.8$) and richness bins ($15 \leq N < 20$, $20 \leq N < 25$, etc.). Bins that do not meet these thresholds are merged with neighboring bins. The resulting 32 stacked groups are shown in Fig.~\ref{fig:stacking}.

For each stacked group, we compute weighted averages of bolometric luminosity, optical richness, and weak-lensing (WL) mass. The WL mass signal-to-noise ratio is used as the weight. Optical richness is taken from the CAMIRA catalog construction \citep{2014MNRAS.444..147O, 2018PASJ...70S..20O}, and WL masses are obtained as described in Sect.~\ref{sec:weaklens}.

\subsection{Stacking analysis of X-ray surface brightness}
\label{subsec:stack_sb}

As noted in Sect.~\ref{sec:sample}, approximately 80\% of the clusters lack individual X-ray counterparts in the eFEDS catalog \citep{Liu22}. We therefore separate the sample into X-ray detected and undetected systems to examine whether the latter exhibit systematically different average gas profiles. This division is used solely to ensure a robust characterization of the stacked surface-brightness profiles.

To study the average gas distribution, we divide the clusters into nine subsamples defined by three richness bins ($15 \leq N < 25$, $25 \leq N < 40$, $40 \leq N < 60$) and three redshift bins ($0.1 \leq z < 0.3$, $0.3 \leq z < 0.6$, $0.6 \leq z < 1.2$).

To stack clusters located at different distances consistently, extraction annuli are defined in physical units. X-ray photon counts in the 0.5--2.0~keV band are measured around the CAMIRA center for each cluster, and for each radial bin we convert the observed photon counts to the corresponding rest-frame 0.5--2~keV emissivity before stacking, ensuring that clusters at different redshifts are compared on a consistent basis. Stacked profiles are obtained by summing the exposure-corrected counts across all clusters in each subsample. Background levels are estimated from surrounding regions following the procedure in Sect.~\ref{subsec:stack_lumi}. Both source and background counts are corrected for exposure time and area prior to subtraction to ensure consistent units.

We model the stacked surface-brightness profiles using a one-dimensional $\beta$-model,
\begin{equation}
    I(r) = I_0 \left[ 1 + \left(\frac{r}{r_c}\right)^2 \right]^{-3\beta + 0.5},
\end{equation}
where $I_0$, $r_c$, and $\beta$ denote the central surface brightness, core radius, and slope.

The model is convolved with the instrumental PSF. For each subsample, PSFs are generated using the \texttt{ermldet} tool, averaged, and fitted with a double-Gaussian function, which reproduces the mean PSF to within 10\%. The PSF fitting is restricted to radii within $40''$, beyond which PSF contributions are negligible.

We fit the stacked profiles for both detected and undetected systems using a $\beta$-model with a fixed slope of $\beta = 2/3$ (e.g., \citealt{Ota04}), convolved with the PSF. The two-dimensional convolution is performed in radial form using the Hankel transform (e.g., \citealt{2004JOSAA..21...53G, 2009JOSAA..26.1767B}) as implemented in the Python package \texttt{PyHank}\footnote{\url{https://pyhank.readthedocs.io/en/latest/}}.

Parameter estimation is carried out using MCMC sampling, including an intrinsic fractional-scatter parameter $f_{\mathrm{int}}$ to account for stacking uncertainties and deviations from a single $\beta$-model. We adopt uniform priors: $I_0 \sim \mathcal{U}(10^{-4}, 10) \times 10^{43} \mathrm{erg\, s^{-1}\, kpc^{-2}}$, $r_c \sim \mathcal{U} (10^{-4}, 5)\, \mathrm{Mpc}$, $f_{\mathrm{int}} \sim \mathcal{U} (0, 1)$. Inspection of the MCMC chains confirms that the $\beta$-model adequately describes the stacked profiles for the purpose of evaluating their central flatness.

\section{Weak-lensing mass measurement}\label{sec:weaklens}
We estimate cluster masses within $R_{500}$ from the CAMIRA center using WL calibration. Following \citet{Okabe25}, we use the HSC-Y3 shape catalog, which includes PSF-corrected ellipticities based on the re-Gaussianization method \citep{Hirata03, HSCWL1styr, HSC-3Y-Shape}. We include only galaxies that satisfy the full-color and full-depth criteria, and meet the photometric redshift selection described in \citet{Medezinski18} and \citet{2019PASJ...71...79O}.

The shear profiles are measured in six radial bins over $100$--$3000$\,kpc. We fit these with a spherical Navarro-Frenk-White (NFW) profile \citep{NFW96} characterized by scale radius $r_s$ and central density $\rho_s$, and convert the results to $M_{500}$ and concentration $c_{500}$.

To account for miscentering, we model the observed tangential shear as a combination of centered and miscentered components:
    \begin{eqnarray}
    f_{\rm model}(r)&=&f_{\rm cen}f_{\rm NFW}(r)+ (1-f_{\rm cen})\int^{\infty }_0 P(r')f_{\rm NFW}^{\rm mis}(r,r') dr', \nonumber  \\
\nonumber  \\
    f_{\rm NFW}(r)&=&\Delta \tilde{\Sigma}_+(r) \left(1+\mathcal{L}_z \Sigma(r)\right), \\
    P(r) &=& \frac{r}{\sigma^2}\exp\left[-\frac{r^2}{2\sigma^2}\right].
\end{eqnarray}

Here, $\Delta \tilde{\Sigma}_+$ and $\Sigma(r)$ are the differential and local surface mass densities, respectively, and $\mathcal{L}_z$ accounts for redshift-dependent lensing geometry. The miscentered component $f_{\rm NFW}^{\rm mis}$ is computed using
\begin{equation}
    \Sigma^{\rm mis}(r,r')=\frac{1}{2\pi}\int^{2\pi}_0 d\theta \Sigma(r^2+r'^2-2rr'\cos \theta)
\end{equation}
\citep{2006MNRAS.373.1159Y}. The scale parameter $\sigma$ characterizes the Gaussian distribution of offsets, and $f_{\rm cen}$ is the fraction of correctly centered clusters.

We constrain the model parameters using a log-likelihood function:
\begin{eqnarray}
-2\ln {\mathcal L}&=&\ln(\det(C_{nm})) +  \label{eq:likelihood_wl} \\
 &&\sum_{n,m}(\Delta \Sigma_{+,n} - f_{{\rm model}}(r_n))C_{nm}^{-1} (\Delta
 \Sigma_{+,m} - f_{{\rm model}}(r_m)), \nonumber
\end{eqnarray}
where the covariance matrix $C$ includes contributions from shape noise, photometric redshift errors, and uncorrelated large-scale structure \citep{Schneider98}. 

We use an MCMC analysis to estimate $M_{200}^{\rm WL}$, miscentering fraction $|f_{\rm mis}| = 1 - f_{\rm cen}$, and scale parameter $\sigma$. Flat priors are adopted on $\ln M_{200}^{\rm WL}$ over $\ln(0.05\,[10^{14}M_\odot])$--$\ln(50\,[10^{14}M_\odot])$, and $-1 < f_{\rm mis} < 1$. Permitting $f_{\rm mis}$ to have a negative value prevents an artificial boundary on the posterior distribution at $f_{\rm mis}=0$. A Gaussian prior $\sigma = 0.3475 \pm 0.060$~Mpc is used based on \citet{Okabe25}. The final $M_{200}^{\rm WL}$ is converted to $M_{500}^{\rm WL}$ using the concentration--mass relation from \citet{Bhattacharya13}, assuming shared parameters for both centered and miscentered halos.

For the scaling relation analysis, we also compute ensemble averages of X-ray luminosity and richness for each subsample. To ensure consistent weighting with the WL mass, we apply a lensing-based weight (Fig.~\ref{fig:z_N_w}):
\begin{eqnarray}
    w_{{\rm lens},i}=\frac{\sum_j w_{ij}}{\sum_{i}\sum_j w_{ij}}.
\end{eqnarray}
Here, $w_{ij}$ denotes the weight for the $j$-th background galaxy associated with the $i$-th cluster. If a cluster has no usable background galaxies, its net weight becomes zero, reflecting the fact that it does not contribute to the WL signal. 

The galaxy weight $w_{ij}$ is used to compute the stacked tangential shear profile $\Delta \Sigma_+$. It incorporates both the ellipticity and lensing efficiency weights:
\begin{eqnarray}
w_{ij}=\frac{1}{e_{\rm rms,j}^2+\sigma_{e,j}^2}\langle \Sigma_{{\rm cr}}(z_{l,i},z_{s,j})^{-1}\rangle^2\label{eq:weight},
\end{eqnarray}
where $e_{\rm rms}$ and $\sigma_e$ denote the root mean square of intrinsic ellipticity and the measurement error per component, respectively. The mean inverse critical surface density is computed using the redshift probability distribution, $P(z)$: 
\begin{eqnarray}
 \langle \Sigma_{{\rm cr}}(z_{l,i},z_{s,j})^{-1}\rangle =
  \frac{\int^\infty_{z_{l,i}}\Sigma_{{\rm cr}}^{-1}(z_{l,i},z_{s,j})P(z_{s,j})dz_{s,j}}{\int^\infty_{0}P(z_{s,j})dz_{s,j}}.
\end{eqnarray}
Here, $z_{l}$ and $z_{s}$ denote the cluster and source redshifts. The critical surface mass density is given by $\Sigma_{{\rm cr}}=c^2D_{s}/4\pi G D_{l} D_{ls}$, where $D_s$ and $D_{ls}$ are the angular diameter distances from the observer to the sources and from the lens to the sources, respectively. 

\section{Results} \label{sec:results}
\subsection{Scaling relations} \label{subsec:scaling}
In this subsection, we derive the scaling relations between weak-lensing mass, X-ray luminosity, and optical richness for the CAMIRA cluster sample. We begin by summarizing the theoretical expectations from self-similar models, then describe the regression framework adopted for fitting the relations, and finally present the results for both the full stacked sample and X-ray detected and undetected subsamples.

\subsubsection{Self-similar expectations}
Scaling relations are derived under the assumption that the gravity dominates the formation and evolution of large-scale structures. In this framework, the bolometric X-ray luminosity--mass and richness--mass relations follow the forms:
\begin{equation}
L \propto M^{4/3}, \quad N \propto M.
\end{equation}
Because the critical density evolves with redshift as the Universe expands, the relations must include a redshift dependence, which we express through the Hubble parameter: $E(z) = H(z)/H_0 = [\Omega_{\rm M}(1+z)^3 + \Omega_{\Lambda}]^{0.5}$. The redshift-corrected forms of the scaling relations (see e.g., \citealp{2013SSRv..177..247G,2022hxga.book...65L}) are:
\begin{equation}
L E(z)^{-1} \propto [M E(z)]^{4/3}, \quad N E(z) \propto M E(z), \label{eq:lm}
\end{equation}
which serve as the baseline expectation against which we compare our measurements.

\subsubsection{Regression model and fitting method}
To compare our measurements with these expectations, we adopt a power-law model of the form:
\begin{equation}
\ln\left(\frac{y}{y_p}\right) = a_e + (b + d \ln(ev)) \ln\left(\frac{x}{x_p}\right) + c \ln(ev),
\label{eq:scaling}
\end{equation}
where $x=M^{\mathrm{true}}_{500} E(z)$ is the latent mass variable scaled by redshift, and $y$ represents one of the redshift-corrected observables, $\{ M^{\mathrm{WL}}_{500} E(z), L E(z)^{-1}, N E(z) \}$. The pivot quantities are $ x_p = E(z_{\mathrm{ref}}) \times 10^{14} M_\odot $, and $ y_p = \{ E(z_{\mathrm{ref}}) \times 10^{14} M_\odot, E(z_{\mathrm{ref}})^{-1} \times 10^{42}~\mathrm{erg\,s^{-1}}, E(z_{\mathrm{ref}})\}$, and we define $ev = E(z)/E(z_{\mathrm{ref}})$. Here $a_e$ is the normalization, $b$ is the power-law slope, $c$ and $d$ describe the redshift evolution in the normalization and slope, respectively. The pivot redshift is set to $z_{\mathrm{ref}} = 0.35$, defined as a WL-weighted mean redshift.

We consider two model configurations: 1. Fixed evolution (i.e., $c = d = 0$); 2. Free evolution, allowing $c$ and $d$ to vary.  If the redshift evolution correction is not applied to $ x $ and $ y $, the factors of $ E(z_{\mathrm{ref}})$ in the pivot values are omitted. The intrinsic scatter in luminosity and richness is assumed to follow a log-normal distribution and to be redshift-independent, as the data do not constrain its evolution. The intrinsic correlation coefficient between luminosity and richness, $r_{\rm coeff}$, is fixed to zero. Priors for the WL mass calibration parameters are adopted from simulations (Appendix~\ref{app:WL}).

Each observed quantity is modeled as a Gaussian distribution around its latent true value, and parameters are sampled using Markov Chain Monte Carlo (MCMC). {A detailed mathematical summary of the hierarchical model, likelihood formulation, and prior choices is given in Appendix~\ref{app:fitmethod}. The regression is implemented using the {\tt HiBRECS} hierarchical Bayesian framework \citep{Akino22}, which is designed to handle multivariate scaling relations while accounting for selection effects, regression dilution, and WL mass calibration.
 
Model selection is performed using the Akaike and Bayesian Information Criteria (AIC, BIC), comparing fixed- and free-evolution versions of the scaling relation.

\subsubsection{Results for the full sample}\label{subsubsec:scaling_whole_sample}
The best-fitting parameters of the $L-M$ and $N-M$ relations for the full CAMIRA sample are listed in Table~\ref{tab:scaling2}. Fig.~\ref{fig:scaling2} shows the fitted relations along with the stacked luminosity and richness measurements. 

Model comparison based on AIC and BIC shows that the fixed-evolution model ($c=d=0$) provides the best relative fit when the expected $E(z)$ corrections are applied (Table~\ref{tab:scaling2}, rows~1--4). Without the redshift correction, the free-evolution model (rows~5--8) yields marginally smaller AIC/BIC values; however, the fitted evolution parameters $c$ and $d$ remain consistent with zero owing to their large uncertainties. For this reason, we adopt the $E(z)$-corrected, fixed-evolution model as our fiducial configuration. The resulting best-fit relations are
\begin{equation}
    L E(z)^{-1} \propto [M E(z)]^{1.56^{+0.14}_{-0.12}},\quad 
    N E(z) \propto [M E(z)]^{0.766^{+0.070}_{-0.060}}. \label{eq:scaling_best}
\end{equation}

Posterior distributions for all fitted parameters, including the intrinsic scatter and hyperparameters of the latent mass distribution, are presented in Appendix~\ref{appendix:posterior}. For completeness, the $N-L$ relation derived using the one-dimensional {\tt HiBRECS} routine is summarized in Appendix~\ref{app:N-L}.

\begin{figure*}
    \centering
    \includegraphics[width=0.48\textwidth]{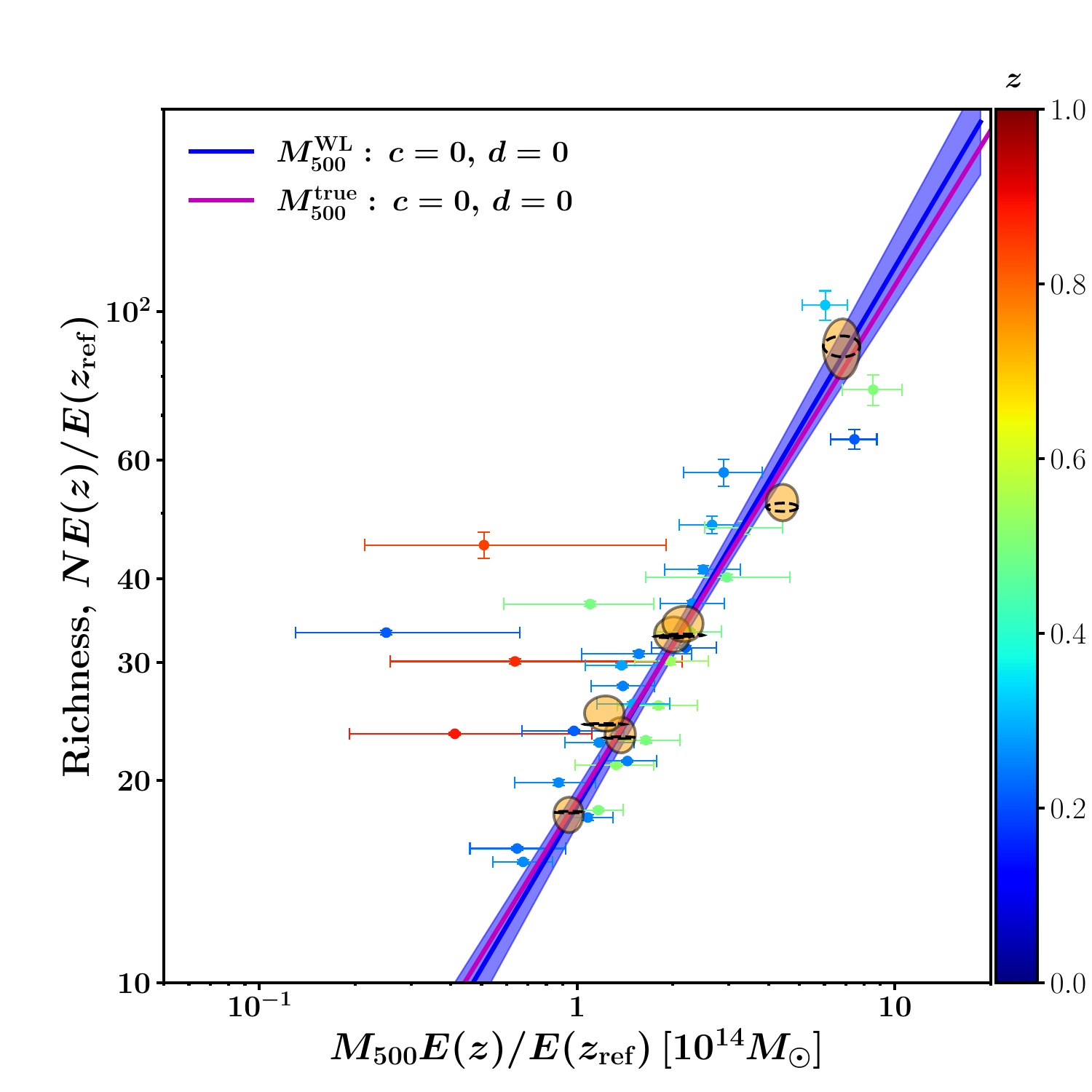}
    \includegraphics[width=0.48\textwidth]{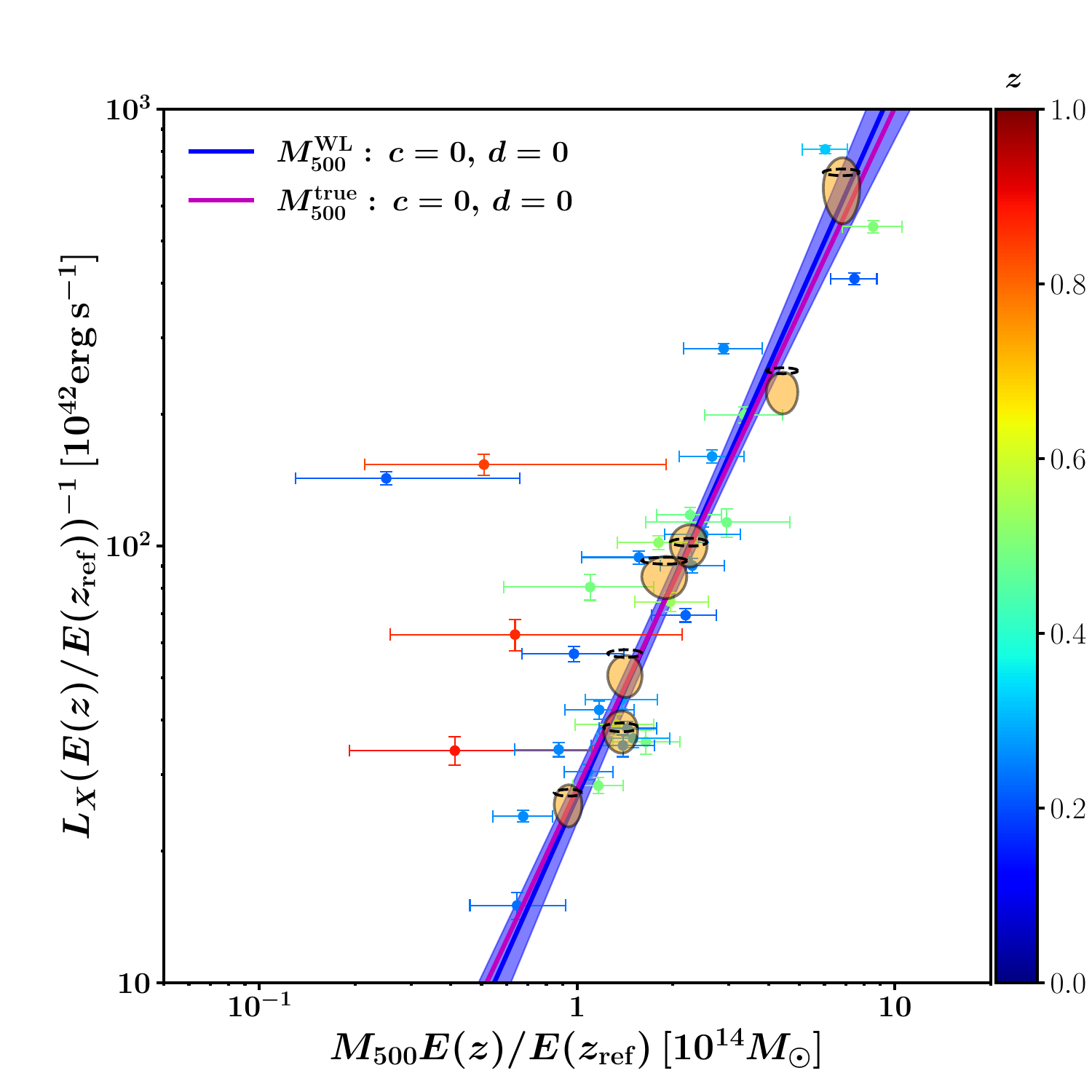}   
    \caption{Scaling relations of the CAMIRA optically selected clusters. Each circle represents a stacked bin described in Sect.~\ref{subsec:stack_lumi}, color-coded by its weighted average redshift. Solid lines show the best-fitting power-law models from the simultaneous fit: blue and magenta correspond to relations with respect to the WL masses and to the true masses, respectively. The shaded blue region indicates the $1\sigma$ uncertainty of the scaling relation with the WL masses. The $1-\sigma$ confidence ellipses indicate the posterior covariance of the stacked quantities, computed using either total uncertainties including intrinsic scatter (orange) or measurement error only (dashed black). The 32 data points are sorted by X-ray luminosity (\textit{left}) or richness (\textit{right}), grouped accordingly, and averaged within the group. }\label{fig:scaling2}
\end{figure*}

\begin{table*}[htb]
\renewcommand{\arraystretch}{1.5}
    \caption{Best-fit parameters for the scaling relations of the CAMIRA clusters.}\label{tab:scaling2}
    \centering
    \begin{tabular}{llllllll} \hline\hline
Relation &  $a$ & $b$ & $c$ & $d$ &$\sigma$ & $\Delta$AIC & $\Delta$BIC\\ \hline
$LE(z)^{-1} - ME(z)$ & $1.444_{-0.059}^{+0.052}$ &  $1.56_{-0.12}^{+0.14}$ & 0 & 0& $0.169_{-0.126}^{+0.073}$ & \multirow{2}{*}{0} &  \multirow{2}{*}{0}\\
$NE(z) - ME(z)$ & $1.272_{-0.030}^{+0.026}$ &  $0.766_{-0.060}^{+0.070}$ & 0& 0& $0.107_{-0.048}^{+0.029}$ \\
$LE(z)^{-1} - ME(z)$ & $1.429_{-0.066}^{+0.054}$ &  $1.61_{-0.13}^{+0.16}$ &  $-1.3_{-2.0}^{+1.7}$ &  $1.0_{-1.9}^{+2.2}$ &  $0.177_{-0.139}^{+0.078}$   &  \multirow{2}{*}{$+10$} &  \multirow{2}{*}{$+18$} \\
$NE(z) - ME(z)$ & $1.261_{-0.033}^{+0.028}$ &  $0.793_{-0.064}^{+0.080}$ & $-0.57_{-1.15}^{+0.90}$ &  $0.48_{-0.97}^{+1.14}$ &  $0.112_{-0.049}^{+0.031}$ \\\hline
$L - M$ & $1.473_{-0.069}^{+0.065}$ &  $1.50_{-0.16}^{+0.16}$ & 0 & 0 & $0.329_{-0.113}^{+0.085}$   &  \multirow{2}{*}{$+20$} & \multirow{2}{*}{$+11$}\\
$N - M$ & $1.258_{-0.037}^{+0.030}$ &  $0.778_{-0.068}^{+0.080}$ & 0 & 0 & $0.134_{-0.068}^{+0.055}$ \\
$L - M$ & $1.430_{-0.060}^{+0.055}$ &  $1.58_{-0.12}^{+0.15}$ & $1.5_{-1.8}^{+1.5}$ &   $-0.1_{-1.3}^{+1.8}$ & $0.195_{-0.146}^{+0.069}$ & \multirow{2}{*}{0} &   \multirow{2}{*}{0}\\
$N - M$ & $1.262_{-0.034}^{+0.028}$ &  $0.779_{-0.063}^{+0.087}$ & $-0.63_{-0.94}^{+0.89}$ &  $-0.04_{-0.78}^{+0.88}$ & $0.107_{-0.060}^{+0.052}$ \\

\hline
\end{tabular}
\tablefoot{Obtained using the two-dimensional {\tt HiBRECS} regression framework. $L$ and $M$ are in units of $10^{42}~{\rm erg\,s^{-1}}$ and $10^{14}M_{\odot}$, respectively. The fitted parameters ($a_e$, $b$, $c$, $d$) are derived using natural logarithms, following Eq.~\ref{eq:scaling}. For compatibility with previous works, we convert $a_e$ to base-10 logarithm in the tables as $a=a_e/(\ln{10})$ so that the scaling relation takes the form: $y=y_p 10^a(x/x_p)^{b+d\ln{ev}}(ev)^c$.
}
\end{table*}

\subsubsection{X-ray detected and undetected sub-groups}\label{subsubsec:scaling_xray_detected}

Clusters with and without X-ray counterparts may exhibit differences in their X-ray scaling relations, either because of variations in their physical states or because of selection effects. Here, we investigate the $L-M$ and $N-M$ scaling relations separately for X-ray detected and undetected CAMIRA clusters in the eFEDS field.

As described in Sect.~\ref{sec:sample}, fewer than 20\% of CAMIRA clusters have a counterpart in the eFEDS X-ray cluster catalog \citep{Liu22}. Following the stacking procedure outlined in Sect.~\ref{subsec:stack_lumi}, we divide the clusters into richness-redshift bins, each containing approximately 500 X-ray counts and at least two clusters. For each bin, we calculate WL-weighted averages of bolometric luminosity, richness, and WL mass. The $L-M$ and $N-M$ relations are then jointly fitted to Eq.~\ref{eq:scaling} (with $c = d = 0$) using the 2D {\tt HiBRECS} method.

The best-fit parameters for the two subgroups are listed in Table~\ref{tab:scaling_xray_detected}. The posterior distributions are shown in Appendix \ref{appendix:posterior} (Figs.~\ref{fig:posterior-distribution-2} and \ref{fig:posterior-distribution-3}). As shown in Table~\ref{tab:scaling_xray_detected} and Fig.~\ref{fig:scaling_xray_detected}, the normalization of the $L-M$ relation is consistent between the X-ray detected and undetected samples. The slope for the detected clusters ($b = 1.80^{+0.34}_{-0.26}$) is steeper than that of the undetected clusters ($b = 1.17^{+0.26}_{-0.21}$), although the difference is modest (approximately the $2\sigma$ level).

While the normalization of the X-ray-undetected clusters is slightly higher, the $N-M$ slopes of the two subgroups agree within uncertainties. This concordance suggests that richness remains a stable mass proxy that is relatively insensitive to selection effects, in contrast to X-ray luminosity. 

Our analysis includes a correction for the richness-threshold selection in the CAMIRA catalog as part of the scaling-relation fitting procedure, and this correction is applied uniformly to both subgroups. Consequently, the observed differences between the two sets of scaling relations likely reflect intrinsic physical differences. However, given the modest significance of the difference in the $L-M$ slopes, this distinction might diminish if additional sources of uncertainty, such as the intrinsic spread of stacked luminosities, were taken into account.

\begin{figure*}
     \centering
    \includegraphics[width=0.48\textwidth]{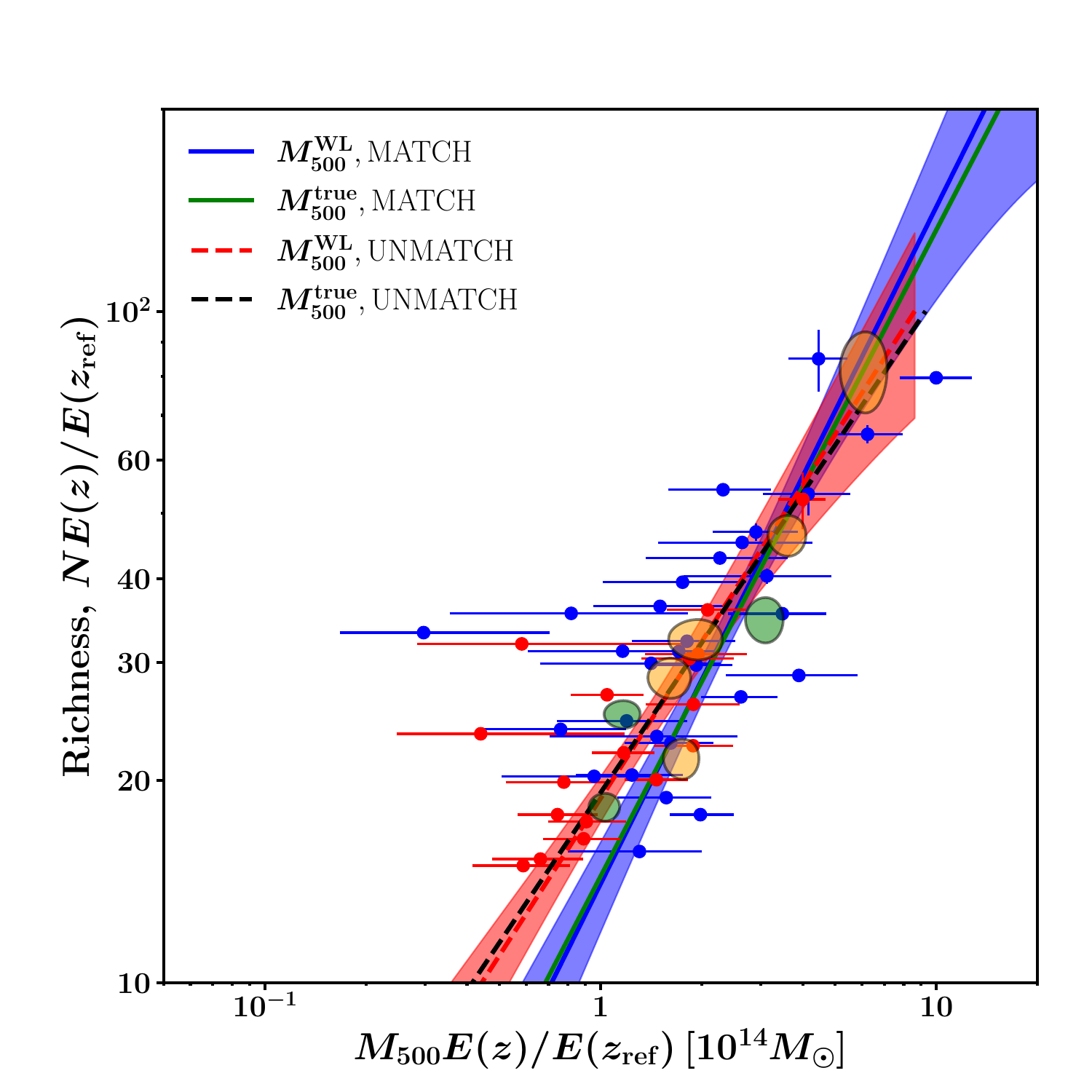}
    \includegraphics[width=0.48\textwidth]{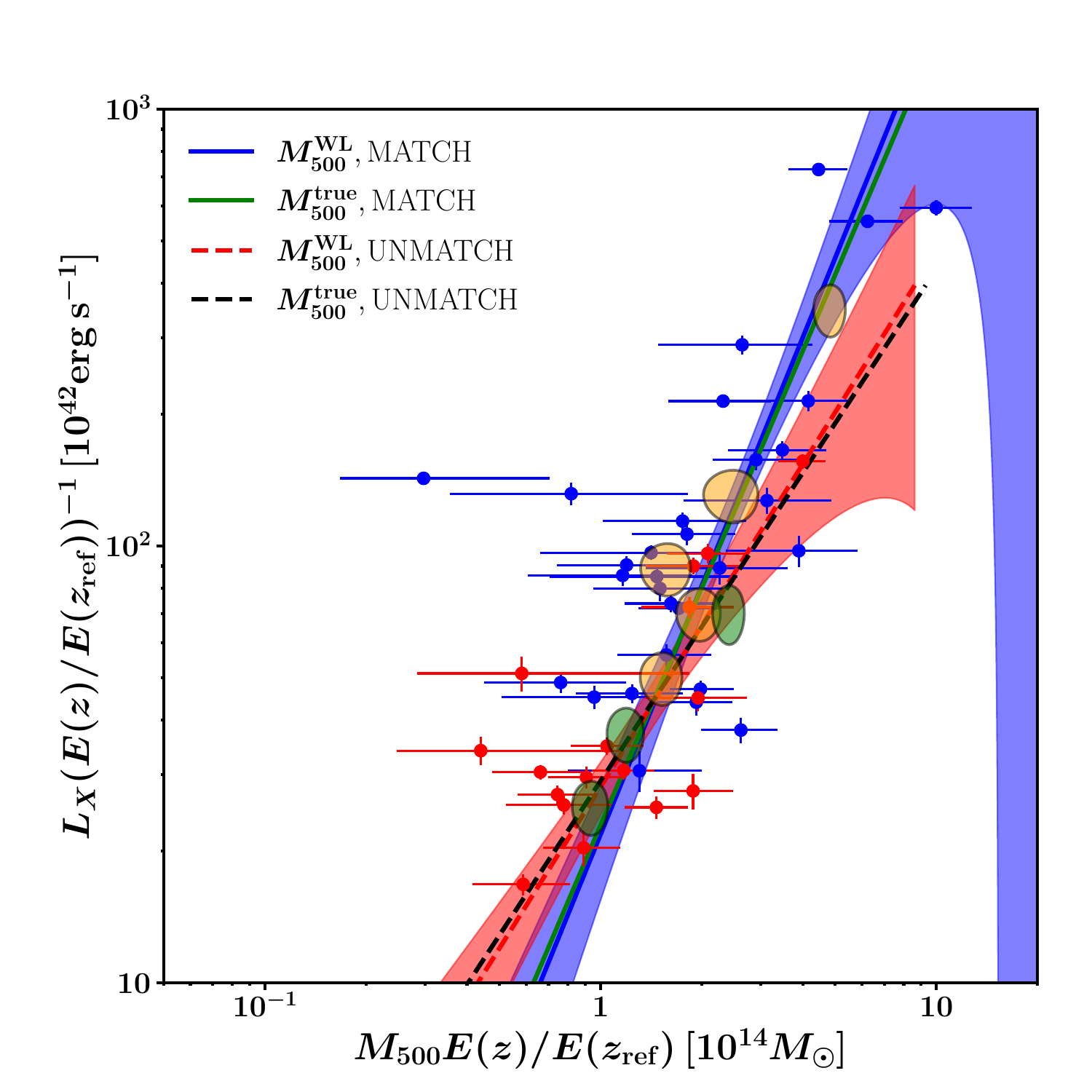}   
    \caption{Scaling relations of the CAMIRA clusters. Circles show stacked bins with approximately 500 X-ray counts. Blue (MATCH) and red (UNMATCH) markers indicate clusters with and without X-ray counterparts in eFEDS, respectively. Solid and dashed lines are best-fit power laws for detected and undetected samples. Best-fit relations are shown with respect to WL (blue/red) and true (green/black) masses. Shaded areas represent $1\sigma$ uncertainties for WL-based fits. The filled orange (X-ray detected) and green (X-ray undetected) ellipses have the same meaning as in Fig. \ref{fig:scaling2}.
    }\label{fig:scaling_xray_detected}
 \end{figure*}

\begin{table*}[htb]
\renewcommand{\arraystretch}{1.5}
    \caption{Best-fit parameters for the X-ray detected and undetected sub-groups.}\label{tab:scaling_xray_detected}
    \centering
    \begin{tabular}{lllll} \hline\hline
Sub-group & Relation & $a$ & $b$ & $\sigma$ \\ \hline
\multirow{2}{*}{X-ray detected} & $LE(z)^{-1} - ME(z)$   & $1.36_{-0.15}^{+0.12}$  & $1.80_{-0.26}^{+0.34}$ &  $0.254_{-0.153}^{+0.080}$   \\
 & $NE(z) - ME(z)$  &$1.158_{-0.079}^{+0.058}$ & $0.96_{-0.12}^{+0.17}$ & $0.138_{-0.075}^{+0.044}$   \\\hline 
\multirow{2}{*}{X-ray undetected} &$LE(z)^{-1} - ME(z)$   & $1.462_{-0.065}^{+0.055}$ & $1.17_{-0.21}^{+0.26}$ &  $0.271_{-0.071}^{+0.072}$ \\
 & $NE(z) - ME(z)$  & $1.283_{-0.039}^{+0.028}$ & $0.74_{-0.10}^{+0.14}$ & $< 0.13$ \\\hline
\end{tabular}
    \tablefoot{
    Same as Table~\ref{tab:scaling2}, but for the X-ray detected and undetected sub-groups.
    }
\end{table*}

\subsection{Surface brightness profiles} 
\label{subsec:surface_brightness}

The best-fitting $\beta$-model parameters are summarized in Table~\ref{table:SB}. In 8 of the 9 richness-redshift bins, X-ray detected clusters show higher central intensities ($I_0$) and smaller core radii ($r_c$) than undetected clusters (Fig.~\ref{fig:SB}), indicating more centrally concentrated surface-brightness profiles. This trend persists when $\beta$ is allowed to vary or when a generalized $\beta$-model with a cusp parameter $\alpha$ is used \citep[e.g.,][]{2002A&A...394..375P,2006ApJ...640..691V, 2016A&A...585A.147A}, and the detected clusters also exhibit smaller residuals relative to the best-fit profiles.

The only exception is the low-redshift ($0.1 \le z < 0.3$), high-richness ($40 \le N < 60$) bin, which contains a single undetected cluster (HSC~J091352--004535). Its central profile is missing because the emission is likely contaminated by a nearby point source, reducing the detection significance. This is unlikely to stem from PSF limitations, as the eROSITA PSF ($\sim$30\arcsec; \citealp{2022A&A...661A...1B}) is still smaller than the cluster extent at these redshifts. No corresponding X-ray source is found in the supplementary eFEDS catalog of compact galaxy groups and clusters \citep{Bulbul22}.

Overall, these results show that X-ray detected clusters possess more concentrated ICM distributions, whereas undetected clusters appear intrinsically fainter or less concentrated, or affected by observational limitations such as contamination from nearby sources.

\begin{figure*}
     \centering
    \includegraphics[width=0.32\textwidth]{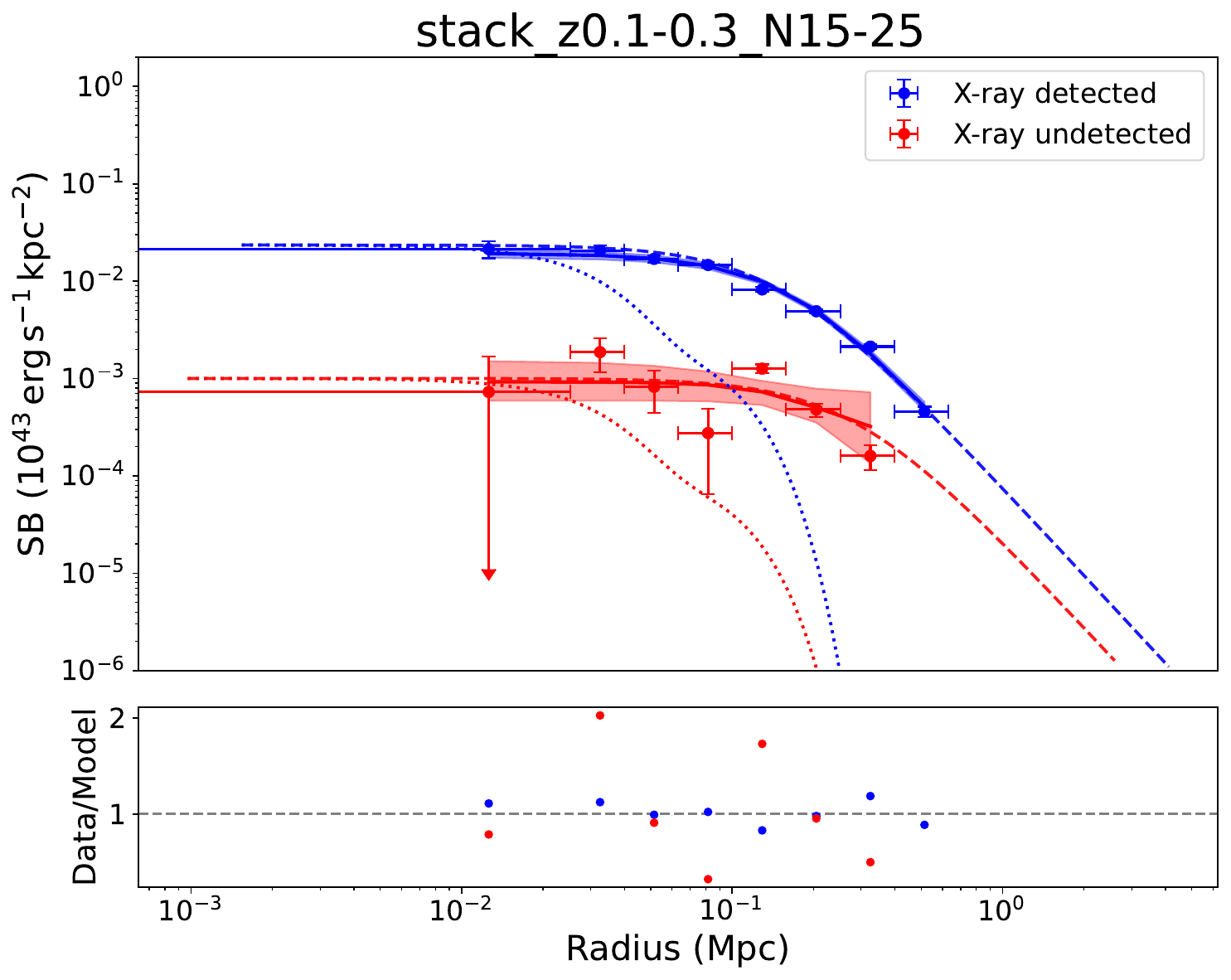}
    \includegraphics[width=0.32\textwidth]{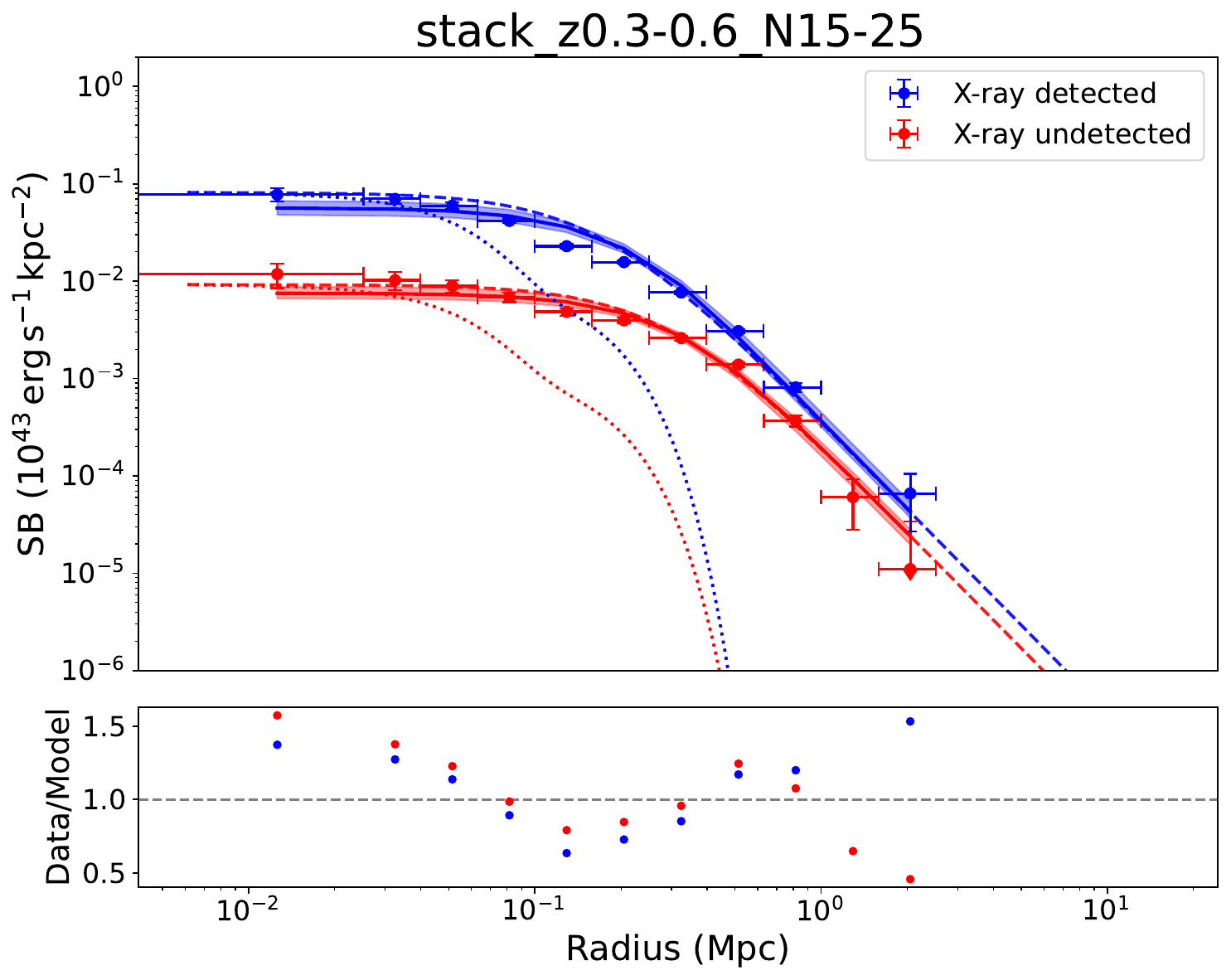}
     \includegraphics[width=0.32\textwidth]{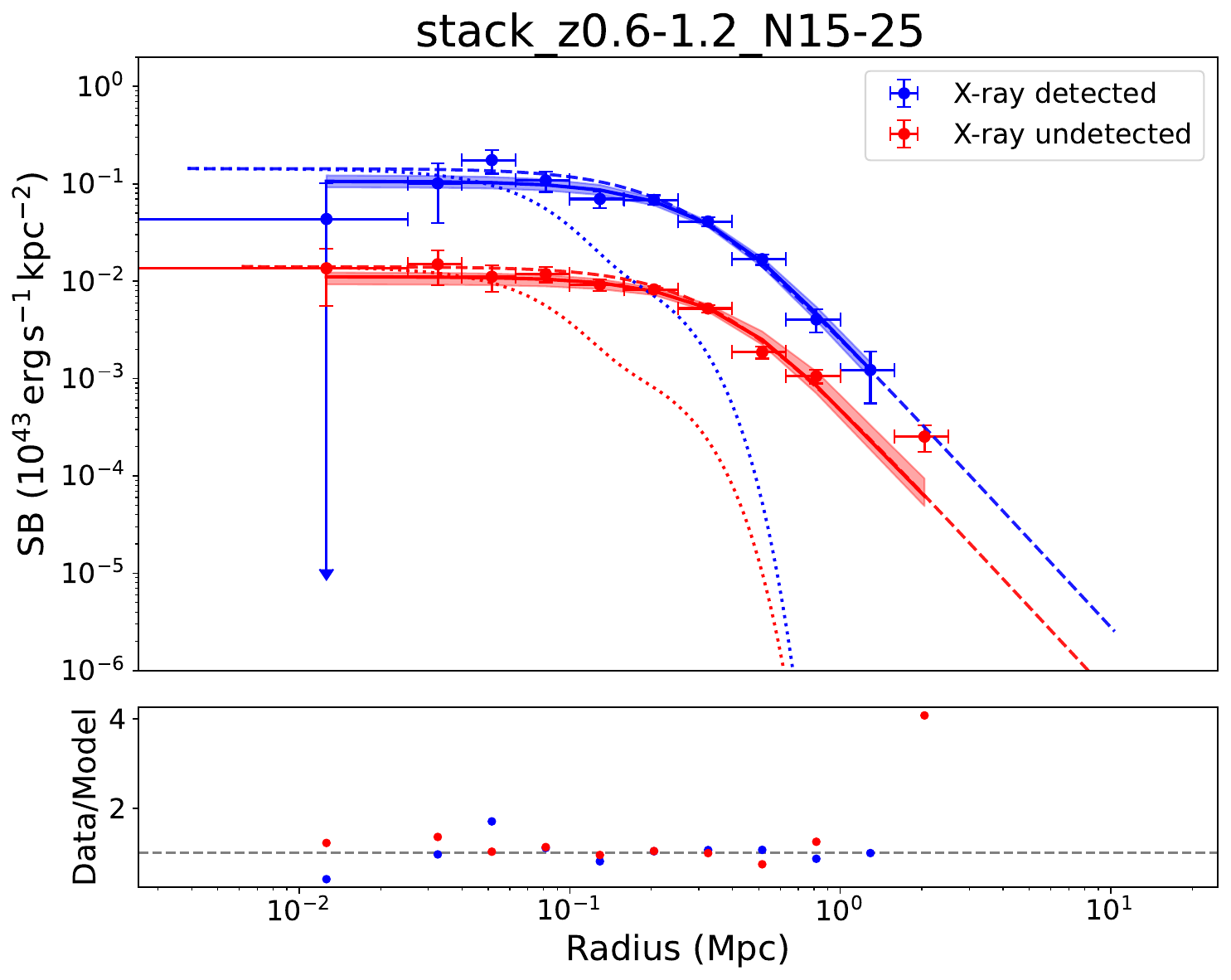}
    
    \includegraphics[width=0.32\textwidth]{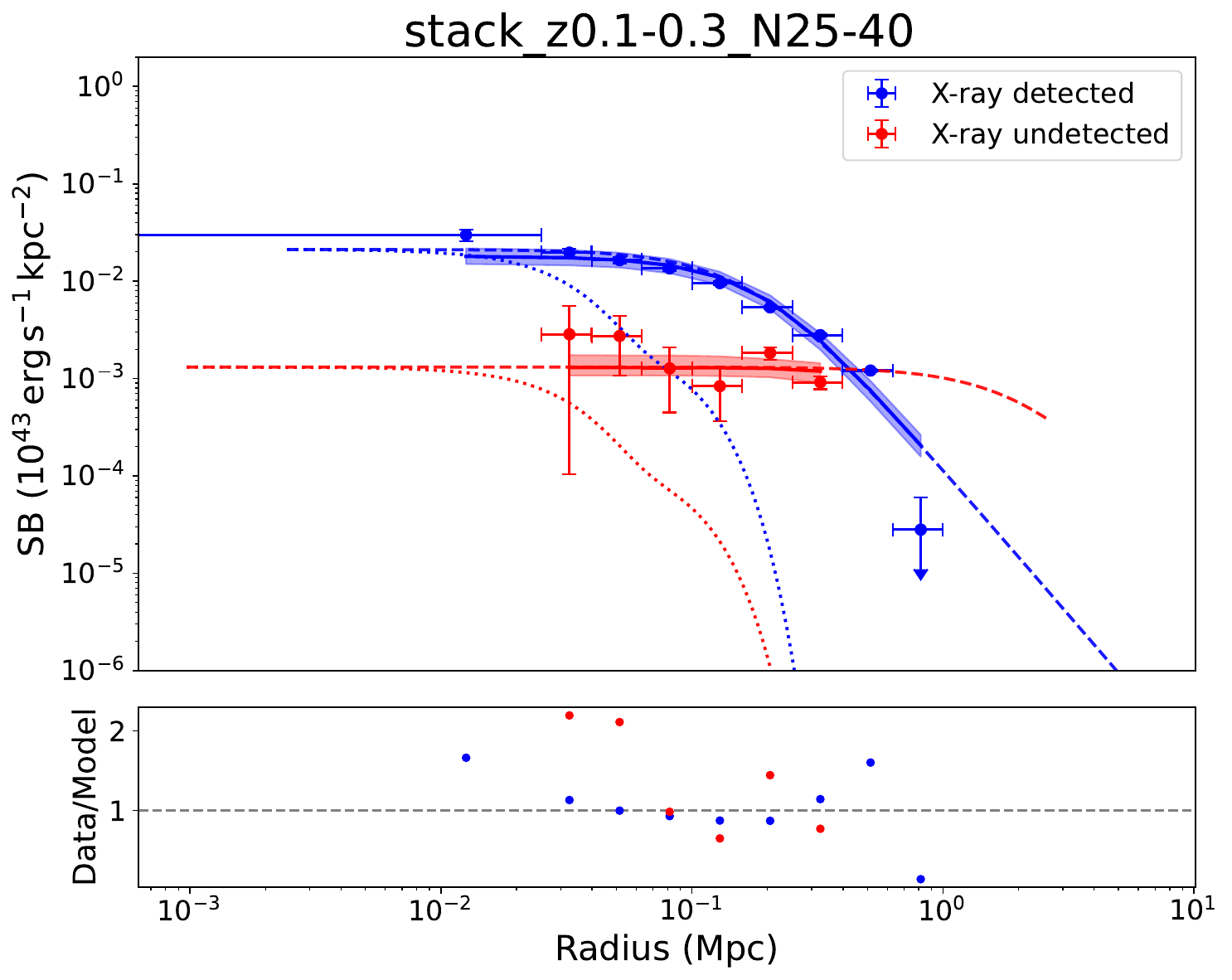}
    \includegraphics[width=0.32\textwidth]{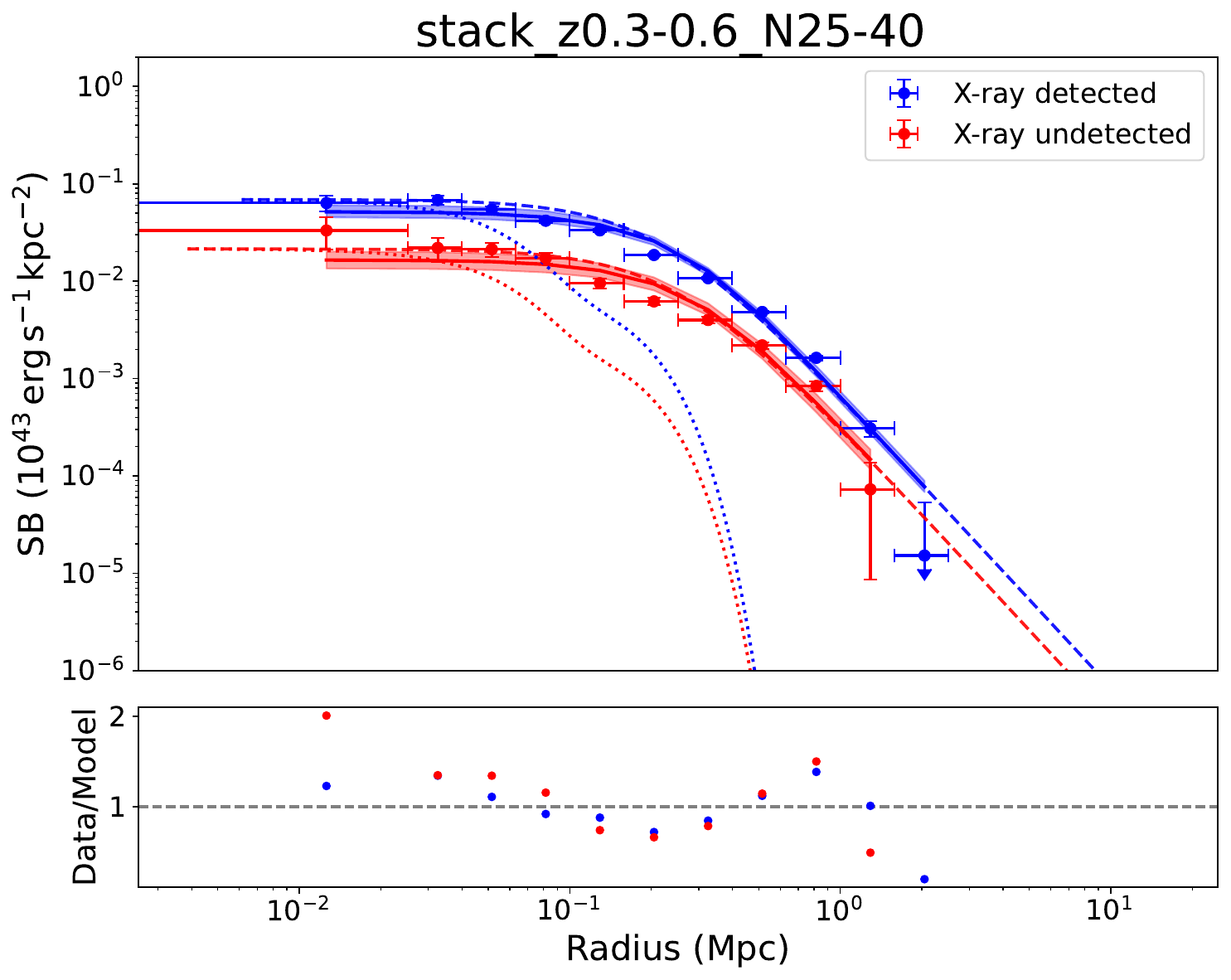}
     \includegraphics[width=0.32\textwidth]{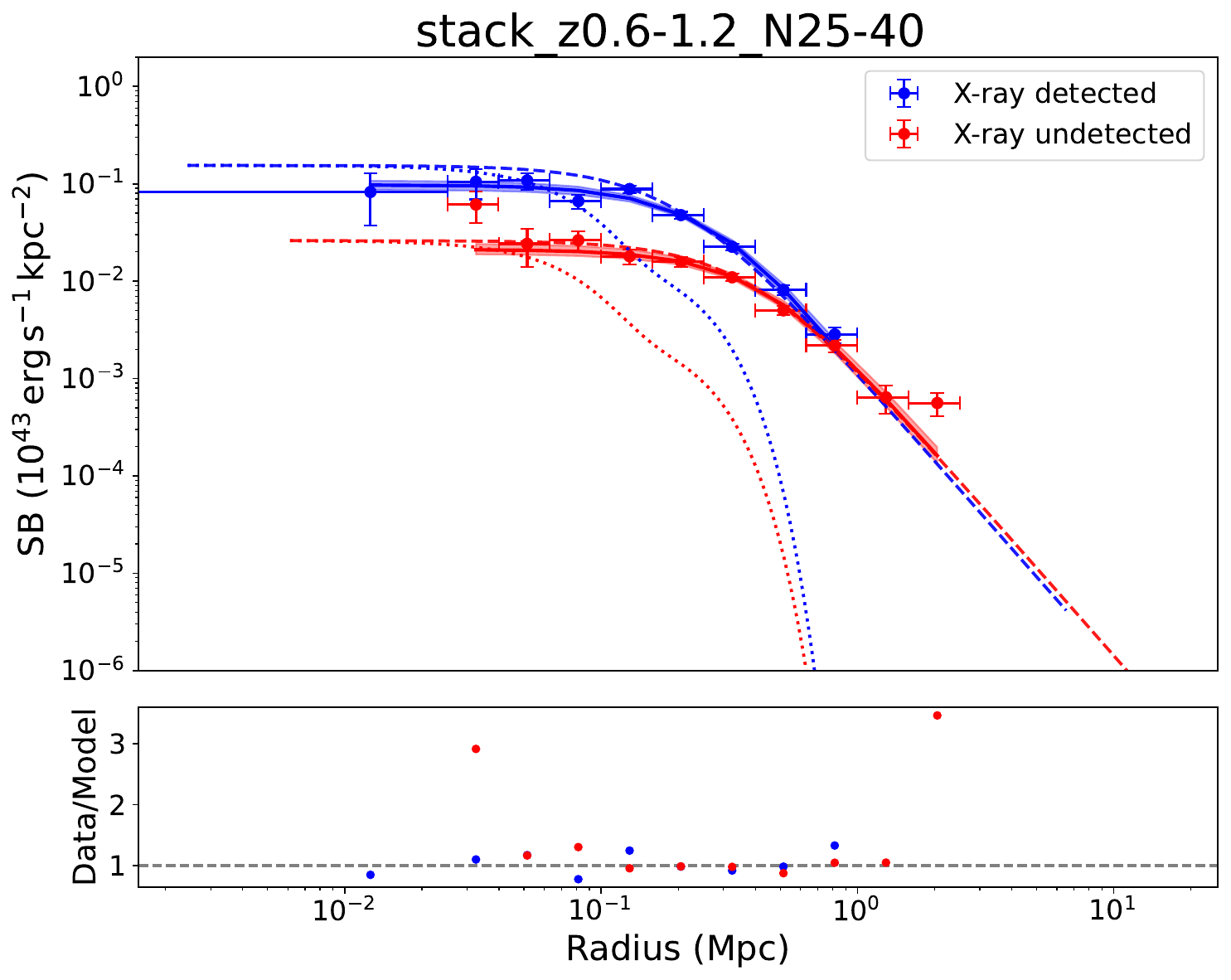}

    \includegraphics[width=0.32\textwidth]{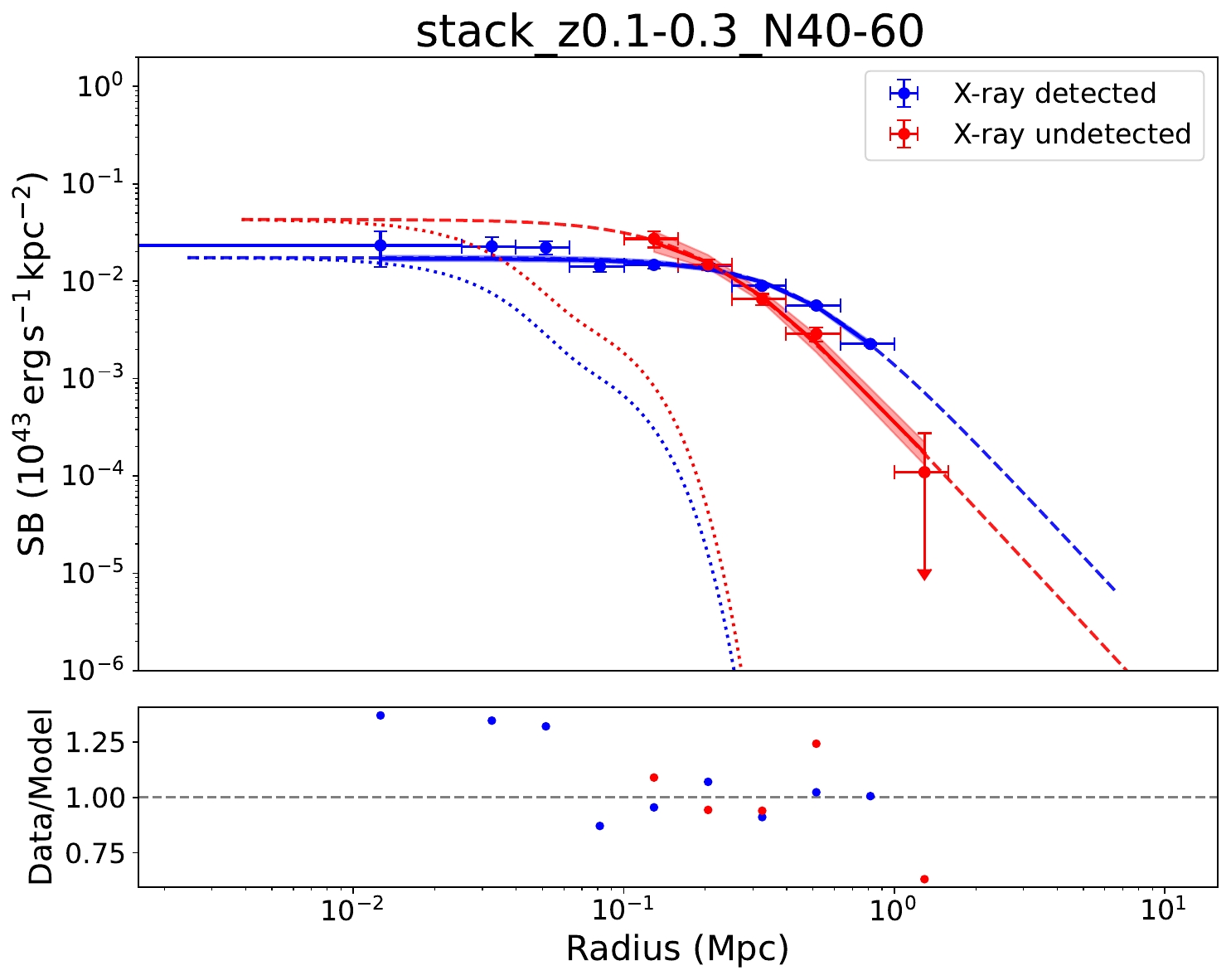}
    \includegraphics[width=0.32\textwidth]{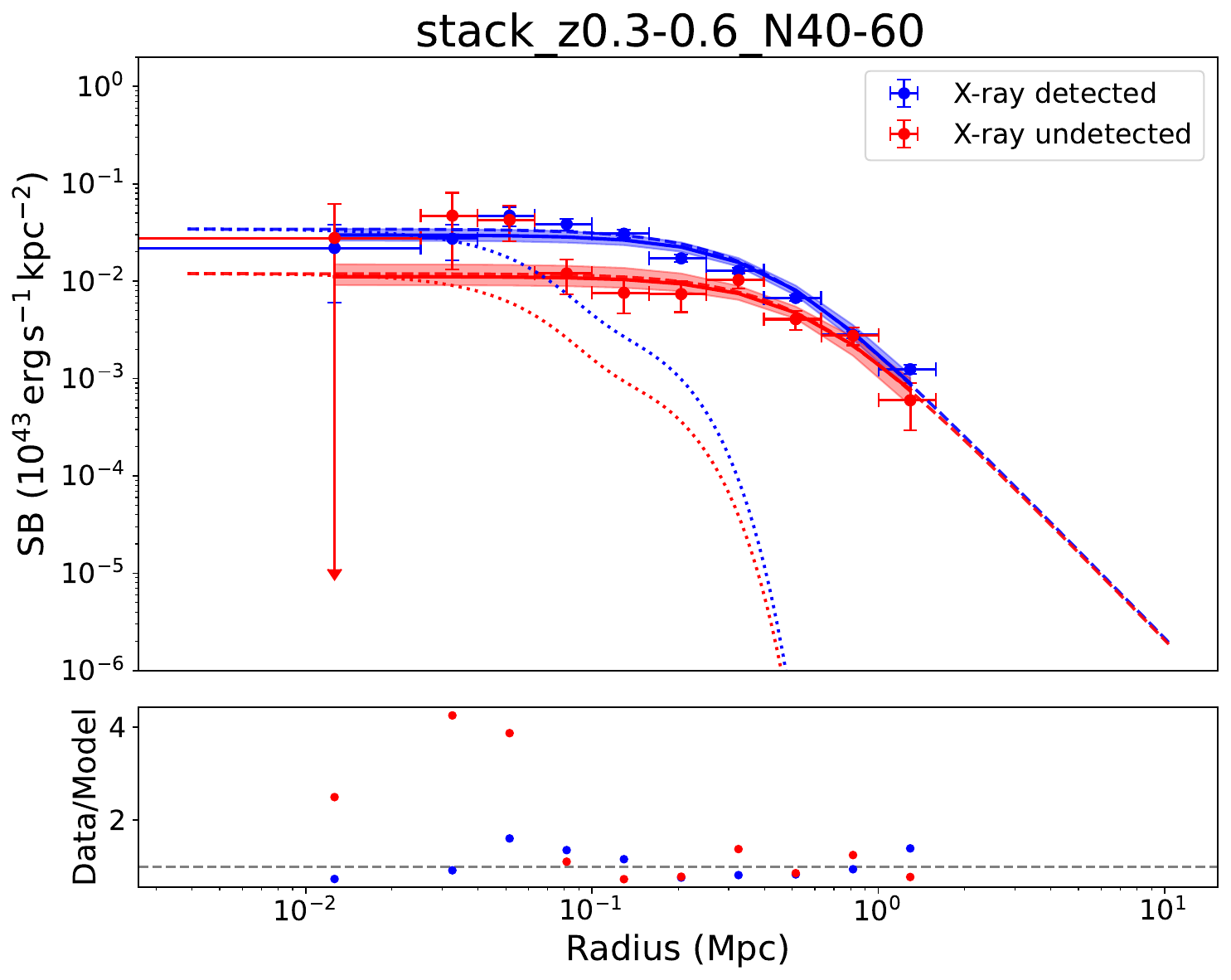}
     \includegraphics[width=0.32\textwidth]{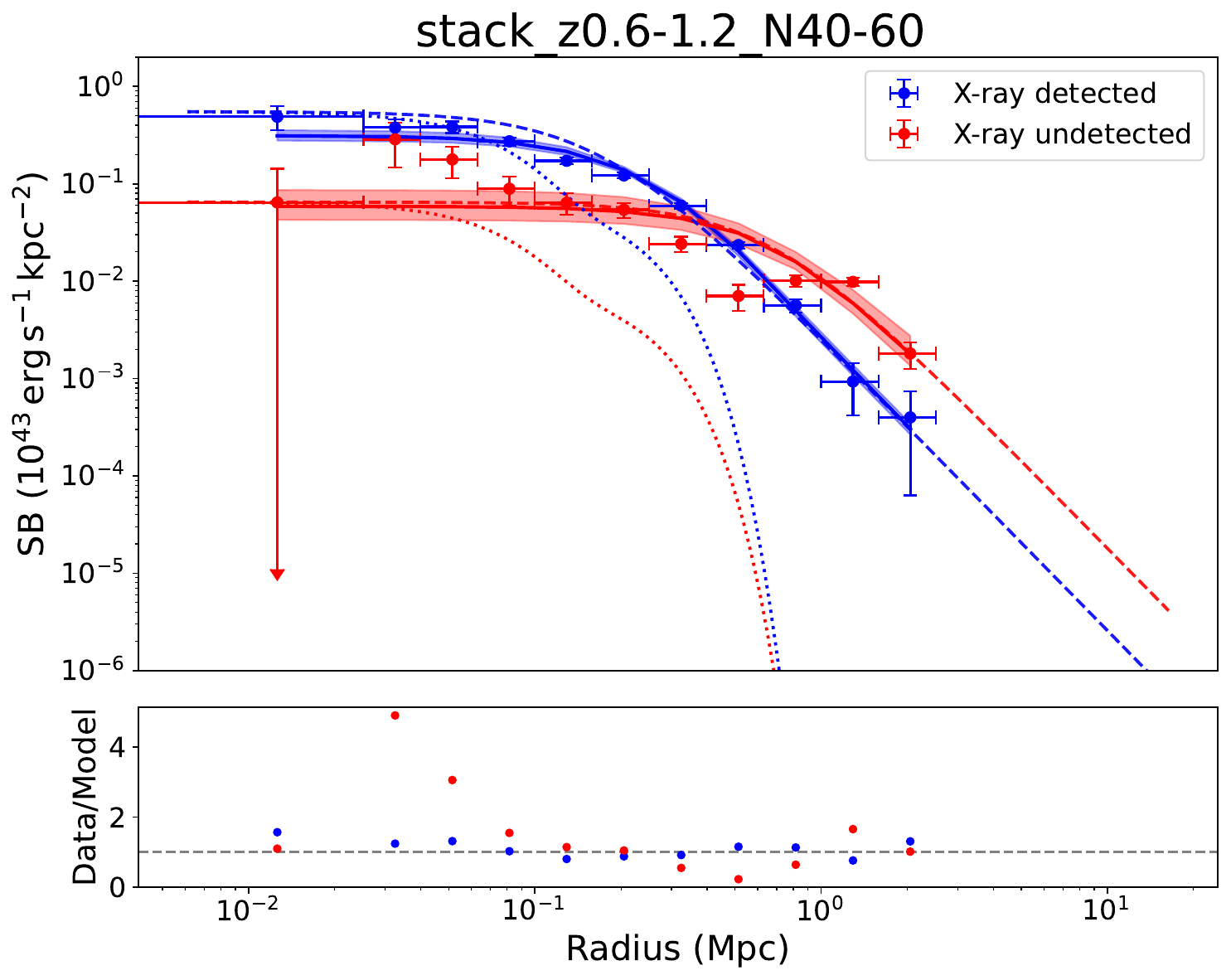}
 \caption{Radial surface brightness profiles of stacked CAMIRA clusters with and without X-ray detections are shown in blue and red circles, respectively. Arrows indicate upper limits. The richness bins are $15 \le N < 25$ (top), $25 \le N < 40$ (middle), and $40 \le N < 60$ (bottom). The left, middle, and right columns correspond to redshift ranges $0.1 \le z < 0.3$, $0.3 \le z < 0.6$, and $0.6 \le z < 1.2$, respectively. Best-fitting PSF-convolved $\beta$-models are plotted as blue and red solid lines, with shaded regions indicating their $1\sigma$ uncertainties. The underlying PSF and intrinsic $\beta$-model components are shown by dotted and dashed lines. The lower panels show the data/model ratios.}
\label{fig:SB}
 \end{figure*}

\begin{table*}[htb]
\renewcommand{\arraystretch}{1.5}
\caption{Best-fit single-$\beta$ model parameters for the surface brightness of CAMIRA clusters.}\label{table:SB}
\centering
\begin{tabular}{c l l l l l}\hline\hline
Subgroup & Richness & Redshift & $I_0$ & $r_c$ & $f_{\mathrm{int}}$ \\ \hline
\multirow{9}{*}{X-ray detected} & \multirow{3}{*}{$15 \leq N < 25$} 
    & $0.1 \leq z < 0.3$ & $ 2.35 \pm 0.26 $ & $ 0.148 \pm 0.010 $ &  $ 0.135 \pm 0.066 $ \\ 
 &  & $0.3 \leq z < 0.6$ &  $ 8.1 \pm 1.6 $ & $ 0.165 \pm 0.018 $ &  $ 0.29 \pm 0.10 $   \\ 
 &  & $0.6 \leq z < 1.2$ &  $ 14.3 \pm 2.5 $ &  $ 0.269 \pm 0.029 $ & $ 0.078 \pm 0.083 $ \\
\cline{2-6} 
 & \multirow{3}{*}{$25 \leq N < 40$} 
    & $0.1 \leq z < 0.3$ & $ 2.12 \pm 0.48 $ &  $ 0.178 \pm 0.025 $ & $ 0.48 \pm 0.18 $ \\ 
 &  & $0.3 \leq z < 0.6$ & $ 6.9 \pm 1.0 $ & $ 0.214 \pm 0.017 $ & $ 0.249 \pm 0.092 $ \\ 
 &  & $0.6 \leq z < 1.2$ & $ 15.5 \pm 2.6 $ & $ 0.195 \pm 0.021 $ & $ 0.129 \pm 0.088 $ \\
\cline{2-6} 
 & \multirow{3}{*}{$40 \leq N < 60$} 
     & $0.1 \leq z < 0.3$ & $ 1.75 \pm 0.12 $ & $ 0.472 \pm 0.026 $ & $ 0.080 \pm 0.061 $  \\ 
 &  & $0.3 \leq z < 0.6$ &  $ 3.43 \pm 0.58 $ & $ 0.398 \pm 0.041 $ & $ 0.26 \pm 0.11 $ \\ 
 &  & $0.6 \leq z < 1.2$ & $ 55 \pm 10 $ & $ 0.167 \pm 0.016 $ &  $ 0.168 \pm 0.084 $ \\ 
\cline{1-6} 

\multirow{9}{*}{Undetected} & \multirow{3}{*}{$15 \leq N < 25$} 
    & $0.1 \leq z < 0.3$ & $ 0.100 \pm 0.064 $ & $ 0.3 \pm 1.3 $ & $ 0.61 \pm 0.21 $ \\ 
 &  & $0.3 \leq z < 0.6$ & $ 0.92 \pm 0.15 $ & $ 0.285 \pm 0.029 $ & $ 0.198 \pm 0.091 $  \\ 
 &  & $0.6 \leq z < 1.2$ & $ 1.41 \pm 0.26 $ & $ 0.342 \pm 0.060 $ &  $ 0.15 \pm 0.12 $  \\ 
\cline{2-6}  
  & \multirow{3}{*}{$25 \leq N < 40$} 
    & $0.1 \leq z < 0.3$ &  $ 0.131 \pm 0.037 $ & $ 2.3 \pm 1.8 $ &  $ 0.40 \pm 0.24 $ \\ 
 &  & $0.3 \leq z < 0.6$ &  $ 2.15 \pm 0.56 $ &  $ 0.248 \pm 0.033 $ & $ 0.36 \pm 0.15 $  \\ 
 &  & $0.6 \leq z < 1.2$ & $ 2.61 \pm 0.39 $ & $ 0.380 \pm 0.039 $ & $ 0.075 \pm 0.087 $ \\ 
\cline{2-6} 
 & \multirow{3}{*}{$40 \leq N < 60$}
    & $0.1 \leq z < 0.3$ & $ 4.2 \pm 2.0 $ & $ 0.207 \pm 0.044 $ & $ 0.17 \pm 0.20 $ \\ 
 &  & $0.3 \leq z < 0.6$ & $ 1.19 \pm 0.35 $ & $ 0.56 \pm 0.12 $ & $ 0.22 \pm 0.20 $  \\
 &  & $0.6 \leq z < 1.2$ & $ 6.5 \pm 2.4 $ & $ 0.65 \pm 0.15 $ & $ 0.54 \pm 0.17 $ \\ 
\cline{1-6} 
\end{tabular}
\tablefoot{
Summary of the best-fit single-$\beta$ model, convolved with the eROSITA PSF, for the surface brightness of CAMIRA clusters with and without X-ray detections in the eFEDS field.
Central brightness $I_0$ and core radius $r_c$ are in units of $10^{41}~\mathrm{erg}\,\mathrm{s}^{-1}\,\mathrm{kpc}^{-2}$ and Mpc, respectively.
}
\end{table*}

\section{Discussion}\label{sec:discussion}
In this section, we interpret our results. We first compare the derived scaling relations with previous work, then examine differences between X-ray detected and undetected clusters, and finally discuss the implications of the stacked surface-brightness profiles.

\subsection{Scaling relations} \label{subsec:discussion_scaling}

\subsubsection{The full sample}
\label{subsubsec:discuss_scaling_whole_sample}

We compare the $L-M$ and $N-M$ results derived for the full stacked sample (Sect.~\ref{subsubsec:scaling_whole_sample}; Eq.~\ref{eq:scaling_best}) with previous studies and theoretical expectations. Relative to our earlier analysis of 43 high-richness CAMIRA clusters \citep{2023A&A...669A.110O}, the present work benefits from a larger sample and a redshift-dependent WL mass calibration. At fixed mass, the inferred richness is systematically higher, partly because miscentering corrections---absent in the earlier study---reduce WL mass estimates, especially at low redshift ($0.58\pm0.04$ for $z<0.4$). At high richness ($N\gtrsim40$), the two studies agree well; at lower richness, masses differ slightly but remain consistent within $1\sigma$. The $N-M$ slope also matches other WL-based results \citep[e.g.,][]{2019PASJ...71..107M,2020MNRAS.495..428C}.

Table~\ref{tab:scaling-comparison} and Fig.~\ref{fig:comparison} summarize literature constraints on the $L-M$ and $N-M$ slopes. Our $L-M$ slope is shallower than those from many X-ray and Sunyaev--Zel’dovich(SZ)-selected samples, which tend to be biased toward massive, relaxed systems, whereas optical selection recovers a broader population, including X-ray--faint clusters. The $N-M$ slope is broadly consistent with previous optical studies, although published values vary with richness definitions.

The theoretical slope of the $L-M$ relation depends on model assumptions. The classical self-similar model \citep{Kaiser86} predicts $L \propto M^{4/3}$, while models including hierarchical structure growth and evolving gas density predict slightly smaller values ($\sim1.1$--$1.2$; \citealt{Fujita19}). Fitting the $L-M$ relation without the $E(z)$ correction (Table~\ref{tab:scaling2}, row~5) yields a slope of $1.49^{+0.15}_{-0.15}$, marginally steeper than both expectations.

Our interpretation adopts the $E(z)$-corrected, fixed-evolution model, identified in Sect.~\ref{subsubsec:scaling_whole_sample} as the preferred description of the data. Allowing the evolution parameters $c$ and $d$ to vary neither improves the fit nor yields significant deviations from zero, indicating no evidence for evolution beyond the self-similar expectation. Within current uncertainties, the results are also consistent with revised baseline models that allow only mild departures from strict self-similarity.

The evolution parameters $c$ and $d$ remain weakly constrained, likely because of the limited sample size and the broad stacking bins, which diminish sensitivity to subtle redshift trends. The literature likewise reports mixed findings on redshift evolution \citep[e.g.,][]{2007ApJ...668..772M,2007MNRAS.382.1289P,2009ApJ...692.1033V,2011A&A...535A...4R,2014A&A...568A..23A}. Larger and more homogeneous samples from future wide-field optical and X-ray surveys will be crucial for strengthening constraints on possible evolution in these relations.

\begin{figure*}
    \centering
    \includegraphics[width=0.48\textwidth]{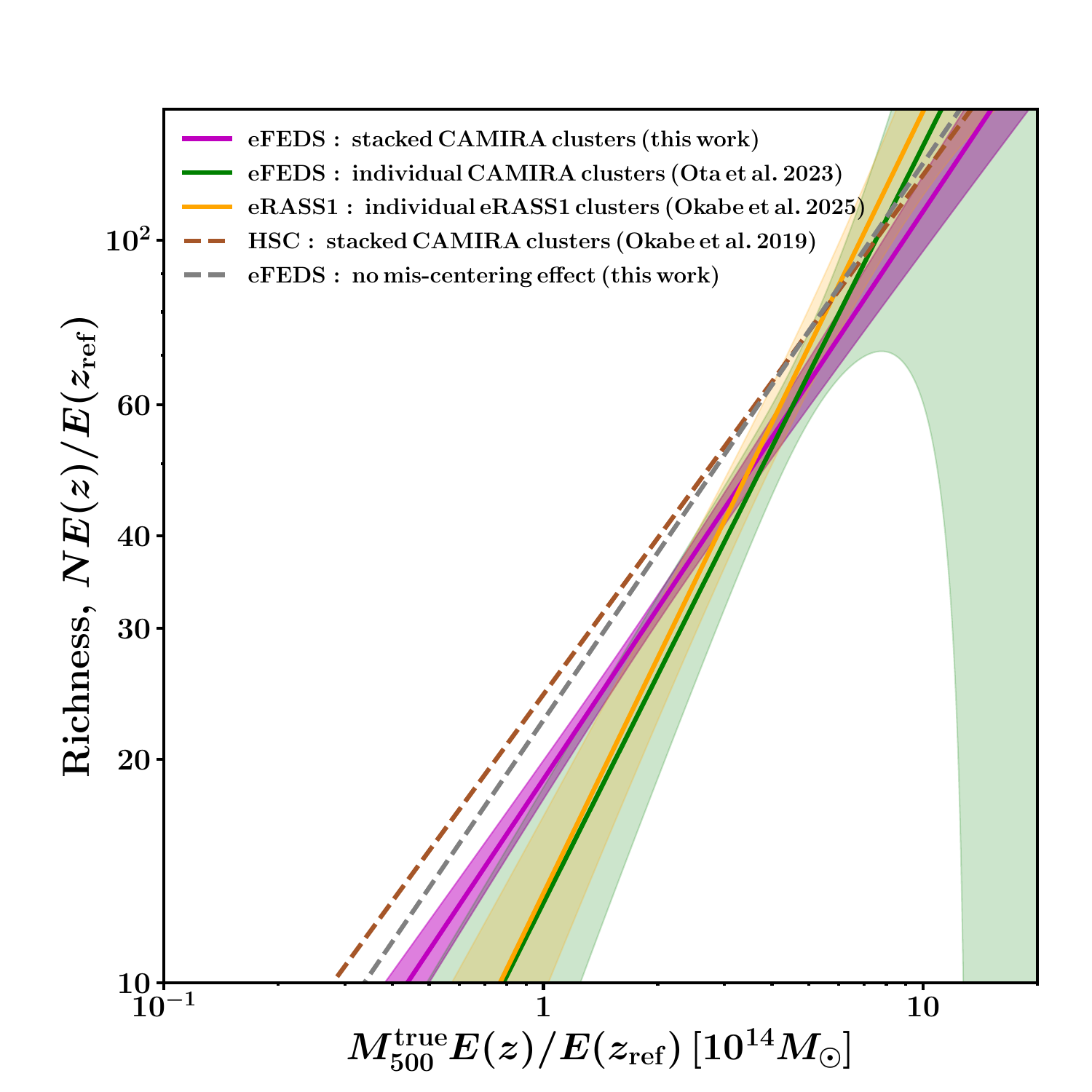}
    \includegraphics[width=0.48\textwidth]{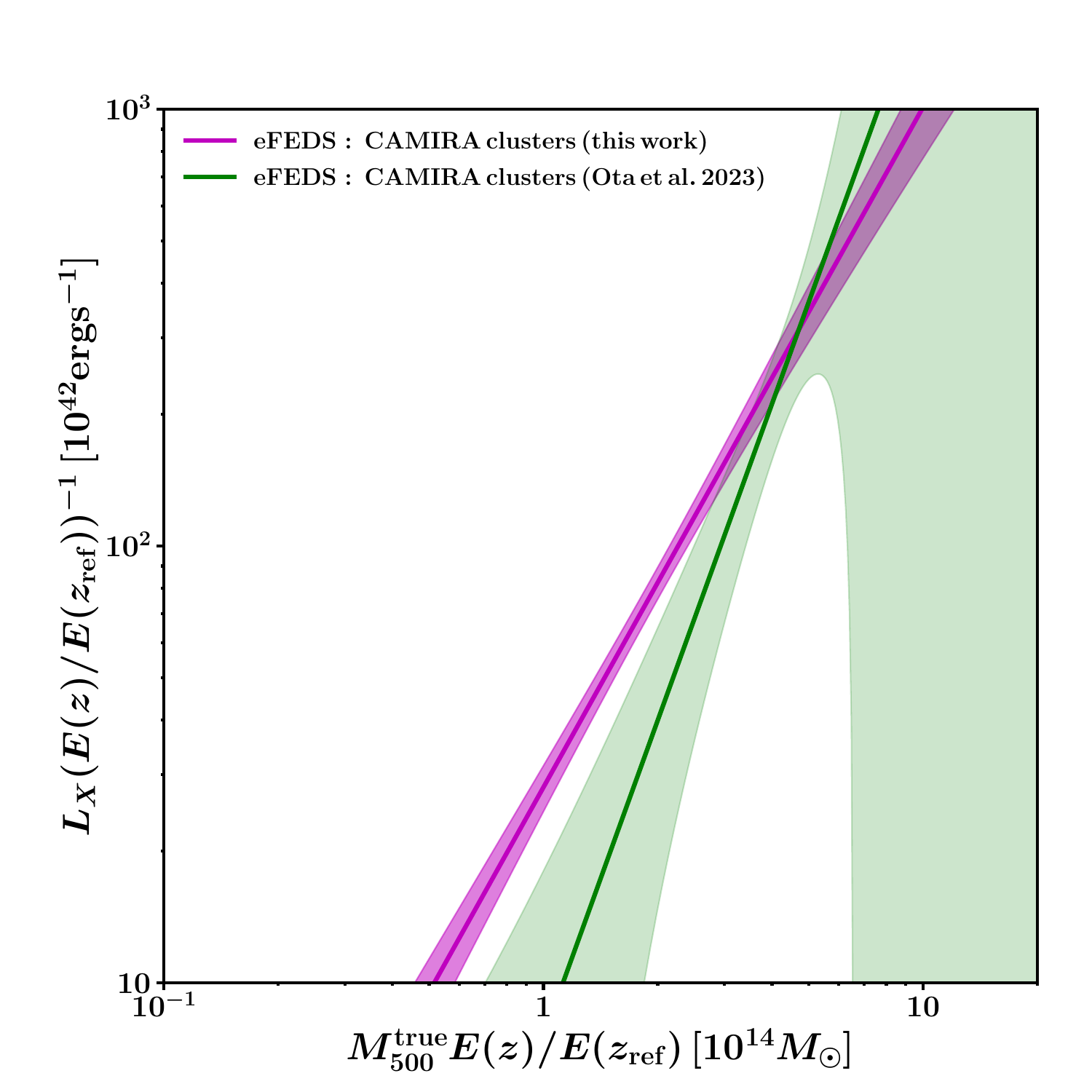}
    \caption{Scaling relations of richness (left) and luminosity (right) with respect to true masses. The magenta solid lines and shaded regions show the best-fit relations and their uncertainties from this work. The green lines represent the results for individual CAMIRA clusters with $N>40$ \citep{2023A&A...669A.110O}. The orange line shows the eRASS1 result \citep{Okabe25}. The brown and gray dashed lines denote the stacked weak-lensing result of \citet{2019PASJ...71...79O} and our relation assuming no miscentering, respectively. All baseline relations are evaluated at $z=0.35$.}\label{fig:scaling3}
\end{figure*}

\begin{table*}[htb]
\caption{Comparison of cluster scaling relations in the literature}\label{tab:scaling-comparison}
\centering
\begin{tabular}{lllllll} 
    \hline
    \hline
    Relation & Best-fit slope & Sample & Cluster $\#$ & $z$ range & Selection method & Reference \\
    \hline
    $L_{\mathrm{bol}} - M$ & $1.56 \pm 0.13$ &  CAMIRA &  32, stacked  & 0.1--1.4 & Optical\tablefootmark{a, b} & This work\\
    & $1.52 \pm 0.34$ & CAMIRA & 38 & 0.16--0.89 & Optical\tablefootmark{a, b} & \multicolumn{1}{l}{\cite{2023A&A...669A.110O}} \\
    & $1.59 \pm 0.14$ & eFEDS  &434 & 0.01--1.3 & X-ray flux limited\tablefootmark{b} & \multicolumn{1}{l}{\cite{2022A&A...661A..11C}} \\
    & 1.56 (fixed) & HSC  & 25 & 0.16--0.65 & Shear selected\tablefootmark{b} & \multicolumn{1}{l}{\cite{2022A&A...661A..14R}} \\
    & $2.15 \pm 0.24$ & SPT &59 & 0.2--1.5 &  SZ\tablefootmark{d} &  \multicolumn{1}{l}{\cite{2019ApJ...871...50B}} \\
    & $1.51 \pm 0.09$ & various & 232 & 0.04--1.46 &  X-ray\tablefootmark{c} &  \multicolumn{1}{l}{\cite{2011A&A...535A...4R}} \\
    & $1.90 \pm 0.11$ & REXCESS & 31 & 0.06--0.17 & X-ray (weakly biased)\tablefootmark{c} &  \multicolumn{1}{l}{\cite{2009A&A...498..361P}} \\
    & $2.33 \pm 0.70$ & ROSAT-based & 37 & 0.14--0.30 & X-ray (unbiased)\tablefootmark{b} &  \multicolumn{1}{l}{\cite{Zhang2008}} \\
    \hline
    $N - M$ & $0.77 \pm 0.07$ & CAMIRA & 32, stacked & 0.1--1.4 & Optical\tablefootmark{a, b} & This work\\ 
    & $0.49 \pm 0.20$ & ROSAT-based & 25 &  0.35--0.65 & X-ray flux limited\tablefootmark{b} &  \multicolumn{1}{l}{\cite{2021MNRAS.502.1494K}} \\
    & $0.92 \pm 0.13$ & CAMIRA & $\sim$\,20, stacked & 0.2--1.1 & Optical\tablefootmark{a, b} &  \multicolumn{1}{l}{\cite{2020MNRAS.495..428C}} \\
    & $0.86 \pm 0.05$ & CAMIRA &$\sim$\,12, stacked & 0.1--1.0 & Optical\tablefootmark{a, b} &  \multicolumn{1}{l}{\cite{2019PASJ...71..107M}} \\
    & $0.72 \pm 0.12$ &  SDSS redMaPPer & $\sim$\,200 & $ <$ 0.35 & Optical\tablefootmark{a, c} &  \multicolumn{1}{l}{\cite{2014ApJ...783...80R}} \\
    & $1.44 \pm 0.27$ & CAMIRA & 5, stacked & 0.1--0.3 & Optical\tablefootmark{a, b} &  \multicolumn{1}{l}{\cite{2014MNRAS.444..147O}} \\
    & $1.3 \pm 0.3$ & CCCP & 23 & 0.15--0.55 & X-ray\tablefootmark{b} &  \multicolumn{1}{l}{\cite{2014A&A...568A..23A}} \\
    \hline
\end{tabular}
\tablefoot{\tablefoottext{a}{sample selected by a red-sequence method.} \tablefoottext{b}{Weak-lensing mass.} \tablefoottext{c}{Hydrostatic mass.}\tablefoottext{d}{SZ effect-based halo mass.}}
\end{table*}

\begin{figure*}
    \centering
    \includegraphics[width=0.48\textwidth]{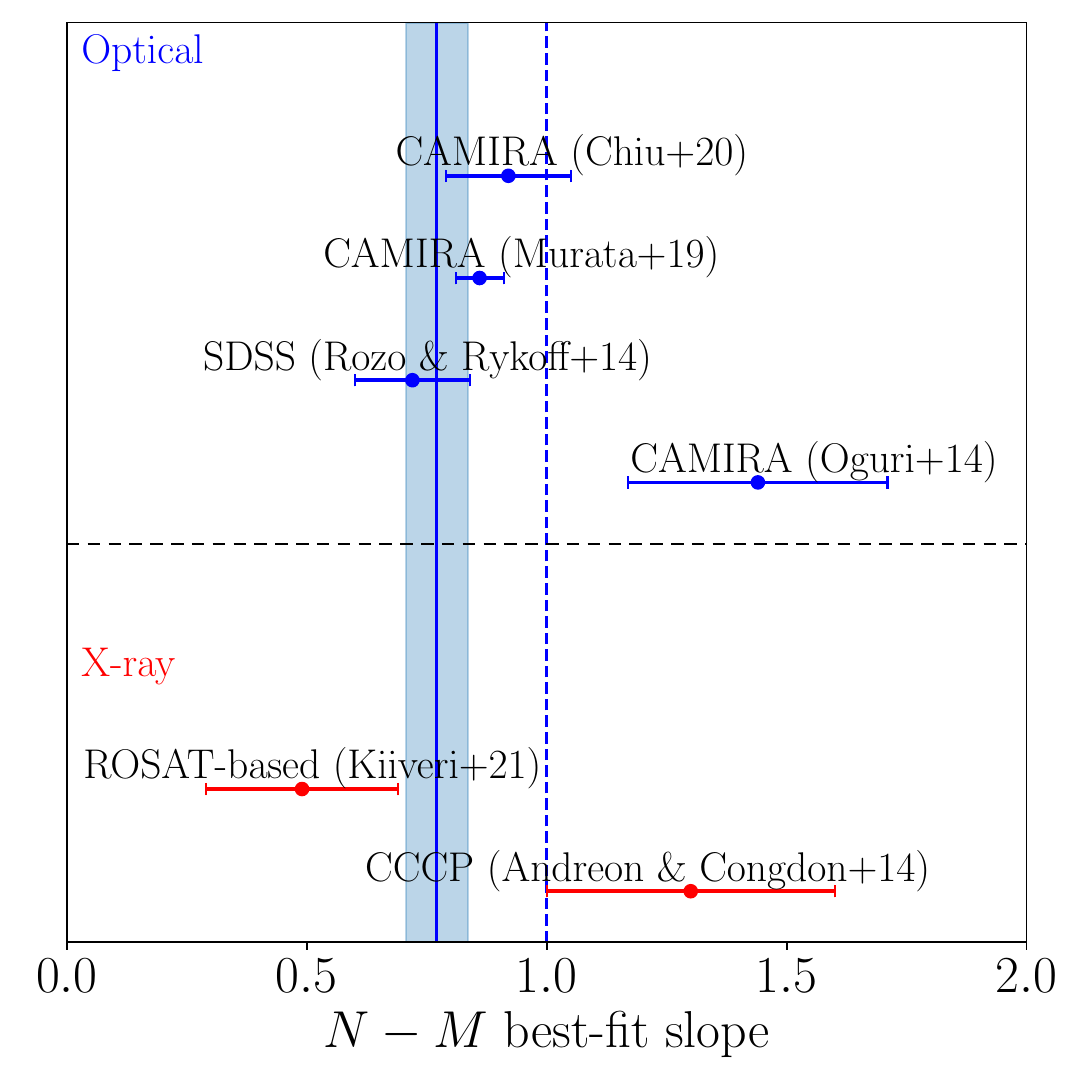}
    \includegraphics[width=0.48\textwidth]{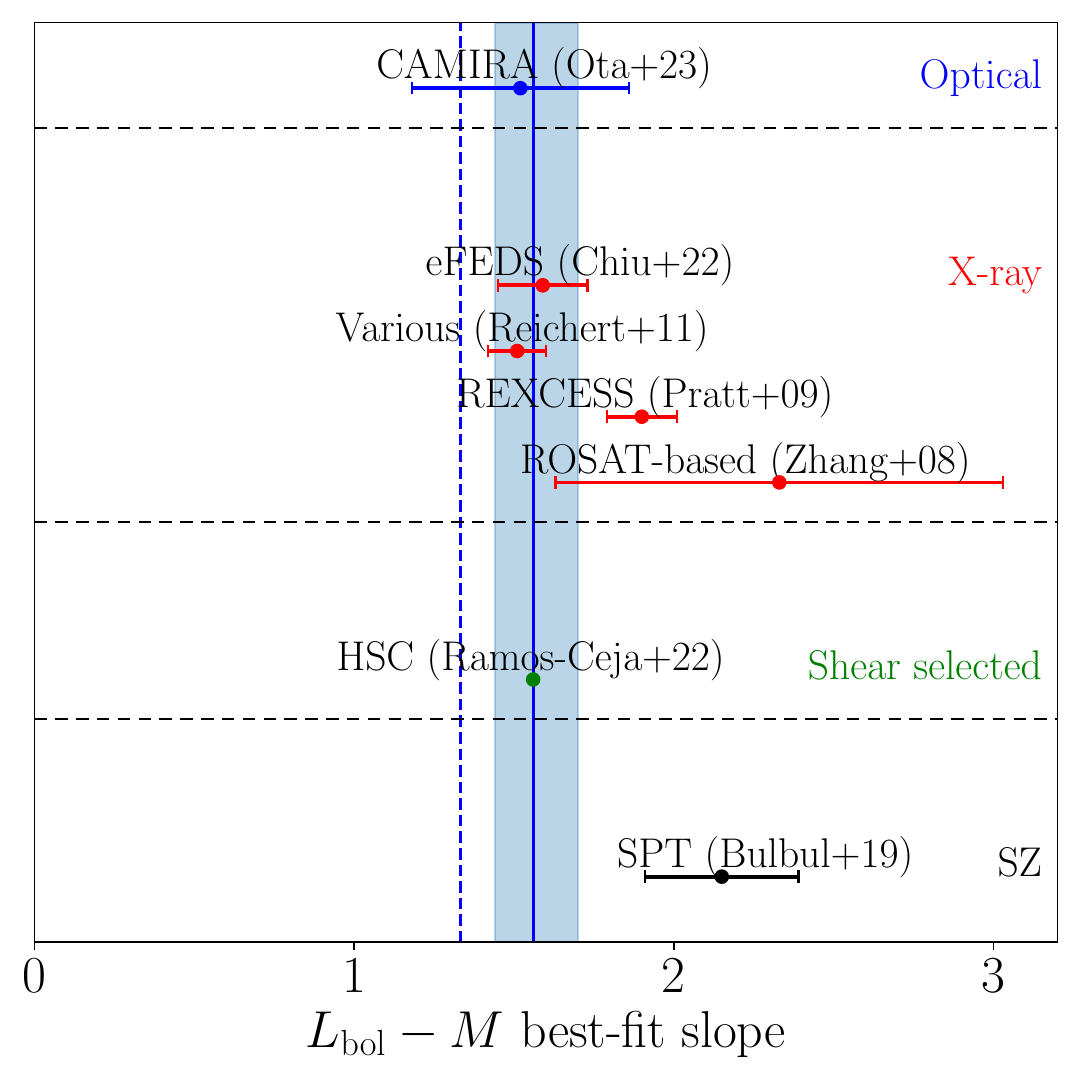}
    \caption{Comparison of the slopes of the $N-M$ (left) and $L-M$ (right) relations.  Solid vertical lines mark the posterior medians from this work, with blue shaded bands indicating the $1\sigma$ credible intervals.  Dashed vertical lines show the slopes expected from self-similar models.  Literature values from Table~\ref{tab:scaling-comparison} are plotted as points with horizontal error bars and are grouped by selection method: optical (blue), X-ray (red), shear-selected (green), and SZ-selected (black). Dashed horizontal lines separate the different selection categories.}\label{fig:comparison}
\end{figure*}

\subsubsection{X-ray detected and undetected sub-groups}
\label{subsubsec:discuss_scaling_xray_detected}
Consistent with previous studies (e.g., \citealt{2017A&A...606A..25A, 2022A&A...661A...7B}), we find that X-ray detected clusters exhibit a marginally steeper $L-M$ slope than undetected ones (Sect.~\ref{subsubsec:scaling_xray_detected}). \citet{2017A&A...606A..25A} showed that X-ray selected samples (e.g., REXCESS) follow a tight $L-M$ relation, while X-ray unbiased samples (e.g., XUCS) show more scatter and shallower slopes. 

This behavior suggests that optically selected clusters with X-ray counterparts resemble the population commonly found in X-ray flux-limited samples and therefore follow a steeper $L-M$ relation. In contrast, clusters without X-ray counterparts tend to be lower-mass systems with a more diffuse ICM. As discussed by \citet{Fujita19}, such systems are expected to form earlier and develop higher gas densities, which can enhance their luminosity and lead to a shallower $L-M$ slope compared with self-similar expectations.

The $N-M$ slopes for the X-ray detected and undetected subsamples, however, are consistent within uncertainties, supporting the view that optical richness remains a robust mass proxy independent of X-ray detectability. This is in line with findings from SDSS redMaPPer and eFEDS-based cluster studies (see Sect.~\ref{subsubsec:discuss_scaling_whole_sample}).

These findings indicate that X-ray-unbiased cluster samples capture a broader diversity of ICM properties. This conclusion aligns with \citet{2016A&A...585A.147A}, who emphasized the importance of optically selected samples for identifying X-ray faint clusters often missed in X-ray or SZ surveys.

\subsection{Surface brightness profiles} 
\label{subsec:discussion_surface_brightness}
X-ray radial profiles provide additional structural insight into differences between X-ray detected and undetected clusters. As shown in Sect.~\ref{subsec:surface_brightness}, CAMIRA clusters with eFEDS counterparts exhibit brighter and more centrally concentrated profiles, whereas undetected clusters display flatter, more extended emission. These trends are consistent with previous observational and simulation-based studies \citep{2020OJAp....3E..13C, 2021MNRAS.503.5624W, 2022A&A...661A...5L, 2024MNRAS.527..895P}.

As in \citet{2021MNRAS.503.5624W}, we consider and rule out miscentering and misclassification of extended sources as point-like. The eROSITA PSF is generally narrower than the typical cluster emission extent (Fig.~\ref{fig:SB}), minimizing its impact. However, contamination by nearby point sources can occasionally reduce the likelihood of detection as an extended source in the eFEDS catalog. A recent eRASS1 study of cluster morphologies \citep{2025A&A...695A.160S} further reported that, at low redshift, eROSITA often misses low-luminosity clusters with flat surface brightness profiles, while at high redshift even concentrated systems may remain undetected. This indicates that many unmatched CAMIRA clusters are genuine systems whose X-ray emission, either intrinsically faint or exhibiting morphologies disfavored by the detection pipeline, falls below the effective sensitivity of current eROSITA catalogs.

The CAMIRA algorithm achieves over 95\% purity for clusters with $\hat{N}_{\mathrm{mem}} \geq 15$ and over 90\% completeness for massive clusters ($M_{200} \geq 10^{14}~M_\odot$), although completeness decreases for low-mass systems at higher redshifts \citep{2018PASJ...70S..20O}. Therefore, some contamination or incompleteness at high redshift cannot be ruled out and may partly explain the mismatch between optical and X-ray samples. Nonetheless, given CAMIRA’s high completeness at low redshift, selection effects alone are unlikely to fully explain the low X-ray detection rate.

The observed differences in surface brightness structure are consistent with the modest ($\sim$2$\sigma$) difference in the $L-M$ slopes between detected and undetected clusters (Sect.~\ref{subsubsec:scaling_xray_detected}), whereas the $N-M$ slope remains consistent. We also verified that excluding low-redshift clusters ($z<0.27$) further reduces the difference between the two subsamples, indicating that the significance of this trend is limited. This implies that, although CAMIRA clusters share similar optical properties, their X-ray characteristics -- including luminosity, radial structure, and detectability -- span a broader range. Optical selection thus recovers a more complete cluster population, including X-ray faint or morphologically diffuse systems.

Recent findings by \citet{2024A&A...686A.284A} also support this interpretation. They find that approximately one-quarter of clusters in a complete sample exhibit low X-ray surface brightness and are missed by both X-ray and SZ surveys (see also \citealp{Mitsuishi2018, Babazaki18, Ota13, Misato2022}). These systems, however, can be successfully identified in deep optical surveys like HSC.

\section{Summary}\label{sec:summary}

Using X-ray stacking analysis, we investigated the X-ray properties of 997 optically selected galaxy clusters from the Subaru HSC CAMIRA catalog in the eFEDS field, covering richness $N > 15$ and redshifts $0.10 < z < 1.34$. Our main results are as follows:

\begin{enumerate}
    \item Scaling relations: We derived bolometric X-ray luminosities and fitted $L-M$ and $N-M$ scaling relations using hierarchical Bayesian regression. The $L-M$ slope is marginally steeper than predicted by self-similar and revised baseline models but is consistent with previous CAMIRA-based studies. The $N-M$ relation broadly agrees with other optical samples. The data do not require any additional redshift evolution beyond the standard self-similar scaling, although the current constraints on evolution remain weak.

    \item X-ray detected vs. undetected: Clusters with eFEDS X-ray counterparts exhibit a steeper $L-M$ slope and more centrally concentrated surface-brightness profiles than undetected systems, indicating systematic differences in their ICM structures. In contrast, the $N-M$ slopes are consistent across the two subgroups, confirming that optical richness is a robust mass proxy regardless of X-ray detectability.

\end{enumerate}

These findings demonstrate that optical cluster selection recovers a broader and more diverse population, indicating X-ray faint and morphologically diffuse systems that are often missed in X-ray surveys. Incorporating such systems extends scaling-relation studies into lower-mass regimes and underscores the importance of coordinated multi-wavelength survey approaches. The eROSITA all-sky data (eRASS:4), in combination with the deep Subaru HSC survey, will enable more precise constraints using substantially larger cluster samples.

\bibliographystyle{aa}
\bibliography{ref}

\appendix

\section{WL mass calibrations} \label{app:WL}
We summarize the calibration procedure for the weak-lensing (WL) mass measurements. Stacking weak-lensing profiles averages over cluster-specific features such as halo ellipticity and substructures, thereby justifying the assumption of spherical symmetry. However, the number density of background galaxies plays a critical role in WL mass measurements. Since observed ellipticities are a combination of intrinsic shapes and the coherent lensing signal, a low background source density may lead to biased mass estimates. In the HSC-SSP survey, the background galaxy number density decreases from $\sim$10 to $\sim$1 arcmin$^{-2}$ as the cluster redshift increases from $z \sim 0.1$ to $z \sim 1$. Consequently, redshift binning can introduce biases in the stacked WL masses.

To quantify the accuracy of our stacked WL mass estimates, we use a mock shape catalog constructed to match our observational setup. We first generate a parent catalog of 10,000 mock clusters assuming spherical NFW halos. Cluster redshifts are uniformly distributed over $z = [0.1, 1.4]$. We include miscentering effects based on the distributions measured in each subsample.

We then generate 2800 subsamples, each having the same number of clusters, redshift range, and approximate mass range as in the real data analysis. Clusters are randomly drawn from the parent catalog without replacement. For each mock subsample, we apply the same WL mass fitting procedure as used for the real data. Mass fitting failed in only about 1\% of subsamples. The resulting mass biases are shown in Fig.~\ref{fig:mbias_vs_M_WLcal}.

To model the WL mass bias, we fit the following relation:
\begin{eqnarray}
    \ln \left(\frac{y_0}{y_{p,0}}\right)=a_{\rm WL}+(b_{\rm WL}+d_{\rm WL}ev(z))\ln \left(\frac{x}{x_p}\right) + c_{\rm WL}ev(z)
\end{eqnarray}
where $x=M_{500}^{\rm true} E(z)$, $y_0=M_{500}^{\rm WL} E(z)$, $x_p=y_{p,0}=10^{14}\,M_\odot E(z_{\rm ref})$, and $ev(z)=\ln (E(z)/E(z_{\rm ref}))$ with $z_{\rm ref}=0.35$. The intrinsic scatter of the WL mass is modeled as:
\begin{eqnarray}
    \sigma_{\rm WL}=\exp\left[\ln \sigma_{\rm WL,0}(1+c_{\ln \sigma} ev(z))\right]. \label{eq:SigmaInt_WL}
\end{eqnarray}
The best-fit parameters are listed in Table~\ref{table:WLcalib}. These WL mass calibrations are used as Gaussian priors in our regression analysis.

\begin{figure}
    \centering
    \includegraphics[width=\linewidth]{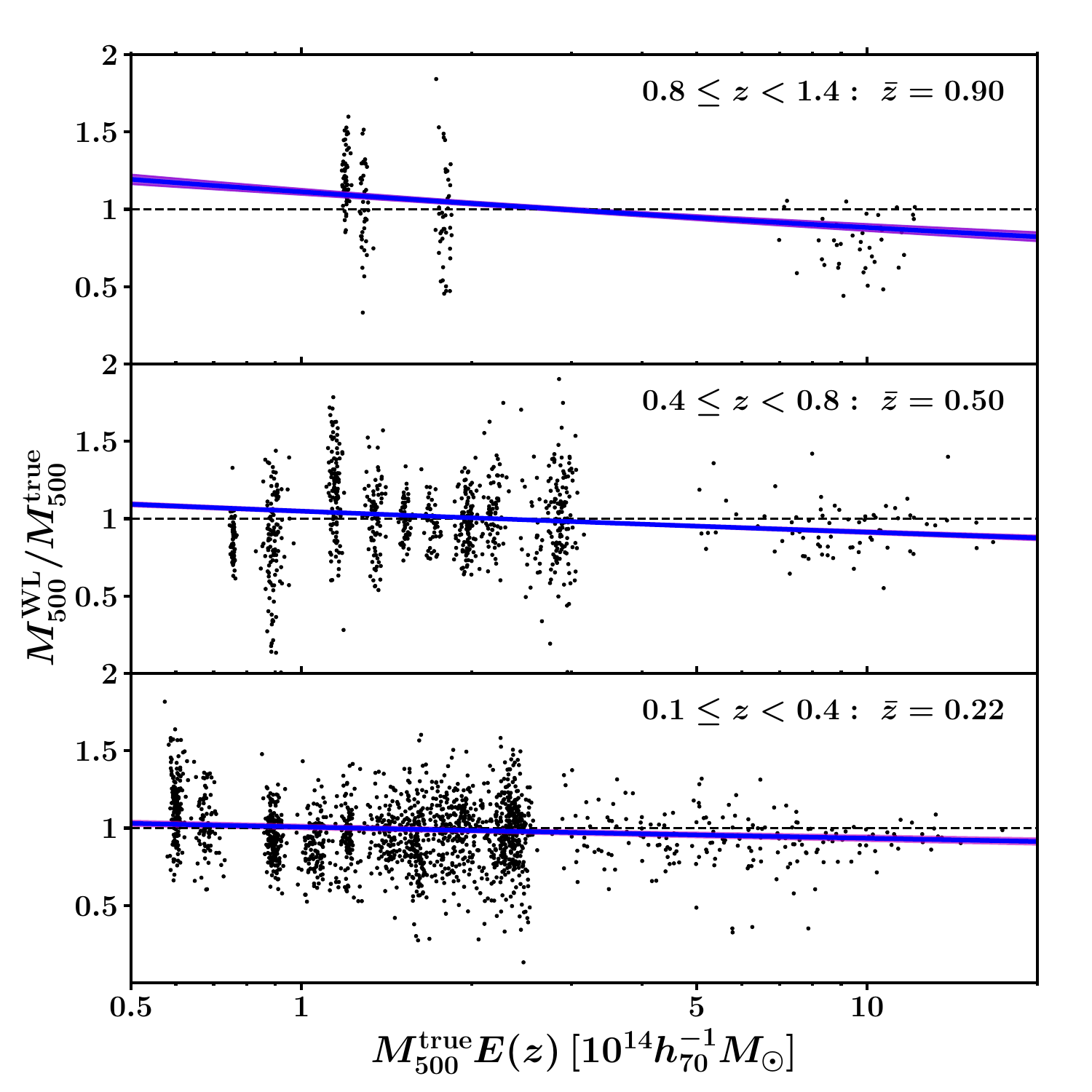}
    \caption{Mass bias $M_{500}^{\rm WL} / M_{500}^{\rm true}$ in three redshift bins. Black points show the bias from stacked mock clusters (without error bars). Blue lines and shaded regions indicate the fitted baseline and its uncertainty. The magenta region represents the baseline including intrinsic scatter.}
    \label{fig:mbias_vs_M_WLcal}
\end{figure}

\begin{table}[h]
\caption{WL mass calibrations for $M_{500}^{\rm WL}$.
} \label{table:WLcalib}
\begin{tabular}{ccc}\hline\hline
$x$ & $M_{500}^{\rm WL}E(z)$ & $M_{500}^{\rm WL}$ \\\hline
 $a_{{\rm WL}}$ & $0.025_{-0.005}^{+0.005}$ & $0.015_{-0.004}^{+0.005}$ \\
 $b_{{\rm WL}}$ & $0.955_{-0.005}^{+0.005}$ & $0.955_{-0.005}^{+0.005}$ \\
 $c_{{\rm WL}}$ & $0.251_{-0.052}^{+0.052}$ & $0.153_{-0.043}^{+0.043}$ \\
 $d_{{\rm WL}}$ & $-0.172_{-0.044}^{+0.044}$ & $-0.180_{-0.044}^{+0.044}$\\
 $\sigma_{{\rm WL0}}$ & $0.006_{-0.003}^{+0.008}$ & $0.007_{-0.003}^{+0.008}$\\
 $c_{{\rm \sigma}}$ & $1.561_{-1.607}^{+1.748}$  & $1.493_{-1.569}^{+1.761}$ \\ \hline
\end{tabular}
\end{table}

\section{Summary of the fitting methodology}\label{app:fitmethod}
We summarize the Bayesian hierarchical framework employed for deriving the scaling relations between the cluster observables. The method follows \citet{2016MNRAS.455.2149S,Akino22} with minor modifications adapted to the present dataset and is implemented using the {\tt HiBRECS} hierarchical Bayesian regression code, which provides a flexible and modular platform for multivariate scaling-relation modeling. This description is provided for completeness and reproducibility.

\subsection*{Regression model}
We assume a log-linear relation between two quantities, $x$ and $y_i$, specified by
\begin{eqnarray} 
    \ln \left(\frac{{y}_i}{{y}_{p,i}}\right)=a_{i} + ({b}_i + {d}_i ev(z))\ln \left(\frac{x}{x_p}\right)+{c}_iev(z) \label{eq:scaling_app}
\end{eqnarray}
where $x=M_{500}^{\rm true}E(z)$, $x_p=10^{14}\,M_\odot E(z_{\rm ref})$, ${\bm y}=\{y_i\}=\{M_{500}^{\rm WL}E(z), NE(z), L_XE(z)^{-1}\}$, and ${\bm y}_p=\{y_{p,i}\}=\{10^{14}\,M_\odot E(z_{\rm ref}), E(z_{\rm ref}), 10^{42}\,{\rm erg\,s^{-1}}E(z_{\rm ref})^{-1}\}$, or  $x=M_{500}^{\rm true}$, $x_p=10^{14}\,M_\odot $, ${\bm y}=\{y_i\}=\{M_{500}^{\rm WL}, N, L_X\}$, and ${\bm y}_p=\{y_i\}=\{10^{14}\,M_\odot, 1, 10^{42}\,{\rm erg\,s^{-1}}\}$, respectively. 
Given the intrinsic covariance in the scaling relation, 
\begin{eqnarray}
    {\bm C}_{\rm int}=\left(
\begin{array}{ccc}
\sigma_{\rm WL}^2 & 0 & 0\\
0 & \sigma_{\ln N}^2 &  r_{\rm coeff}\sigma_{\ln N}\sigma_{\ln L_X} \\
0 &  r_{\rm coeff}\sigma_{\ln N}\sigma_{\ln L_X}  &  \sigma_{\ln L_X}^2 \\
\end{array}
\right), \label{eq:Sigma_int}
\end{eqnarray}
a conditional probability of the scaling relation is described by a trivariate normal distribution, $p({\bm y}|x,{\bm \theta})={\mathcal N}(\ln {\bm y}/{\bm y}_p-{\bm \mu}_x,{\bm C}_{\rm int}$), where ${\bm \mu}_x$ is the right-hand side of the equation (\ref{eq:scaling_app}). Since intrinsic scatter is a non-negative quantity, we treat $\ln \sigma_{\ln N}$ and $\ln \sigma_{\ln L_X}$ as free parameters assuming no redshift dependence. 
When we express the observed quantities as ${\tilde {\bm y}}$, the relationship between $\bm y$ and ${\tilde {\bm y}}$ is expressed by $p({\tilde {\bm y}}|{\bm y})={\mathcal N}(\ln {\tilde {\bm y}}-\ln {\bm y}|{\bm C}_{\rm err})$, where ${\bm C}_{\rm err}$ is the measurement errors. The measurement errors for the $i$-th and $j$-th quantities are given by $C_{\rm err,ij}=\sigma_{{\rm err},ij}^2\delta_{ij}$, where $\delta_{ij}$ is a Kronecker's Delta. Since the measurement errors in the WL masses are asymmetry, we approximate it with
\begin{equation}
    \sigma_{{\rm WL,err}} = \sqrt{0.5\,\left(\sigma_{\rm WL,+}^2 + \sigma_{\rm WL, -}^2\right)} ,
\end{equation}
where $\sigma_{{\rm WL},+}$ and $\sigma_{\rm WL,-}$ are the upper and lower 1$\sigma$ errors, respectively. 

The Bayesian chain rule gives the likelihood function of the trivariate scaling relations (eq. \ref{eq:likelihood}).
Here, the subscript $n$ denotes the $n$-th cluster and $\bm \theta$ is parameters of the scaling relations. 
We introduce the parent population, $p(x|\bm{\theta})$, assuming the Gaussian distribution, $\mathcal{N}(\mu_{\ln x}, \sigma_{\ln x})$, where $\mu_{\ln x}$ and $\sigma_{\ln x}$ are hyper-parameters as a part of the parameters.  
Our optical cluster sample is defined by a richness threshold of $N>15$, which introduces a selection boundary of ${\tilde y}_{{\rm th},1}=15$. This term corrects a selection effect as demonstrated in Fig.~13 of \citet{Akino22}.

In the present application to stacked observables, the scatter term is interpreted as an effective residual scatter of the stacked data points around the mean scaling relation, rather than as the object-by-object intrinsic scatter of individual clusters within each stack.

\begin{table*}[h]
\begin{equation}
  p({\tilde {\bm y}}|\bm \theta)=\frac{\displaystyle \prod_n^{N}\int_{-\infty}^{\infty} \!\!\!d^{3}\ln {\bm y}_n\int_{-\infty}^{\infty}\!\!\!  d\ln x_n\ p({\tilde {\bm y}}_n|{\bm y}_n)p({\bm y}_n|x_n,\bm \theta)p(x_n|\bm{\theta})}{\displaystyle \prod_n^{N} \int_{-\infty}^{\infty}d\ln {\tilde y}_{0,n} \int_{\ln {\tilde y}_{\mathrm{th},1,n}}^{\infty}\!\!\!  d\ln {\tilde y}_{1,n}\!\! \int_{-\infty}^{\infty}\!\!\! d\ln {\tilde y}_2\int_{-\infty}^{\infty} \!\!\! d\ln x_n\ p({\tilde {\bm y}}_n|{\bm y}_n)p({\bm y}_n|x_n,\bm \theta)p(x_n|\bm{\theta})} \label{eq:likelihood}
\end{equation}
\end{table*}

\subsection*{Priors}
We implemented the prior distributions for the WL mass bias parameters, derived from a mock simulation detailed in Appendix \ref{app:WL}. The diagonal components of these distributions are specified in Table \ref{table:WLcalib}. The prior distribution for the slope parameters, $c_i$ and $d_i$, is modeled as a Student's $t_1$ distribution with a single degree of freedom. For the normalization and intrinsic scatter parameters, we utilized flat priors set within the ranges $[-10^4,10^4]$ and $[\ln 10^{-5},\ln 1]$, respectively.  A non-informative prior distribution on the variance of the parent distribution, $\sigma_{\ln x}^2$, follows a scaled inverse $\chi^2$ distribution as a conjugate prior satisfying that posterior distributions have the same probability distribution family as the prior distribution.

\subsection*{Intrinsic correlation coefficient.}
In principle, the intrinsic correlation coefficient $r_{\rm coeff}$ between richness and X-ray luminosity could be treated as a free parameter. We tested this option by allowing $r_{\rm coeff}$ to vary, obtaining $r_{\rm coeff}=-0.22_{-0.49}^{+0.59}$, indicating that the parameter is unconstrained by the current dataset. Furthermore, including this extra degree of freedom slightly worsens the model selection metrics, with $\Delta{\rm AIC} \simeq +2$, $\Delta{\rm BIC} \simeq +5$. Given the lack of constraining power and the AIC/BIC penalty, we adopt $r_{\rm coeff}=0$ in our baseline analysis.

\subsection*{Model comparison and fit assessment}
We evaluate relative model performance using the Akaike Information Criterion (AIC) and Bayesian Information Criterion (BIC). Since these quantities provide relative evidence rather than absolute goodness-of-fit, they are used only for model comparison.

\subsection*{Weak-lensing weights}
For consistency between the scaling-relation analysis and the weak-lensing (WL) mass measurement, we applied the same lensing-based weights used in the stacked tangential shear analysis (see Sect.~\ref{sec:weaklens}). The weight for each cluster, $w_{{\rm lens},i}$, is proportional to the sum of the background-galaxy weights $w_{ij}$ that account for intrinsic shape noise and lensing efficiency, and is normalized by the total weight over all clusters. This scheme ensures that ensemble averages of $L_X$ and $N$ are weighted in a manner consistent with the WL mass calibration. The distribution of $w_{{\rm lens},i}$ for each stack is shown in Fig.~\ref{fig:z_N_w}, illustrating that the adopted weights are well-behaved and without strong outliers.

\begin{figure}
    \centering
    \includegraphics[width=\linewidth]{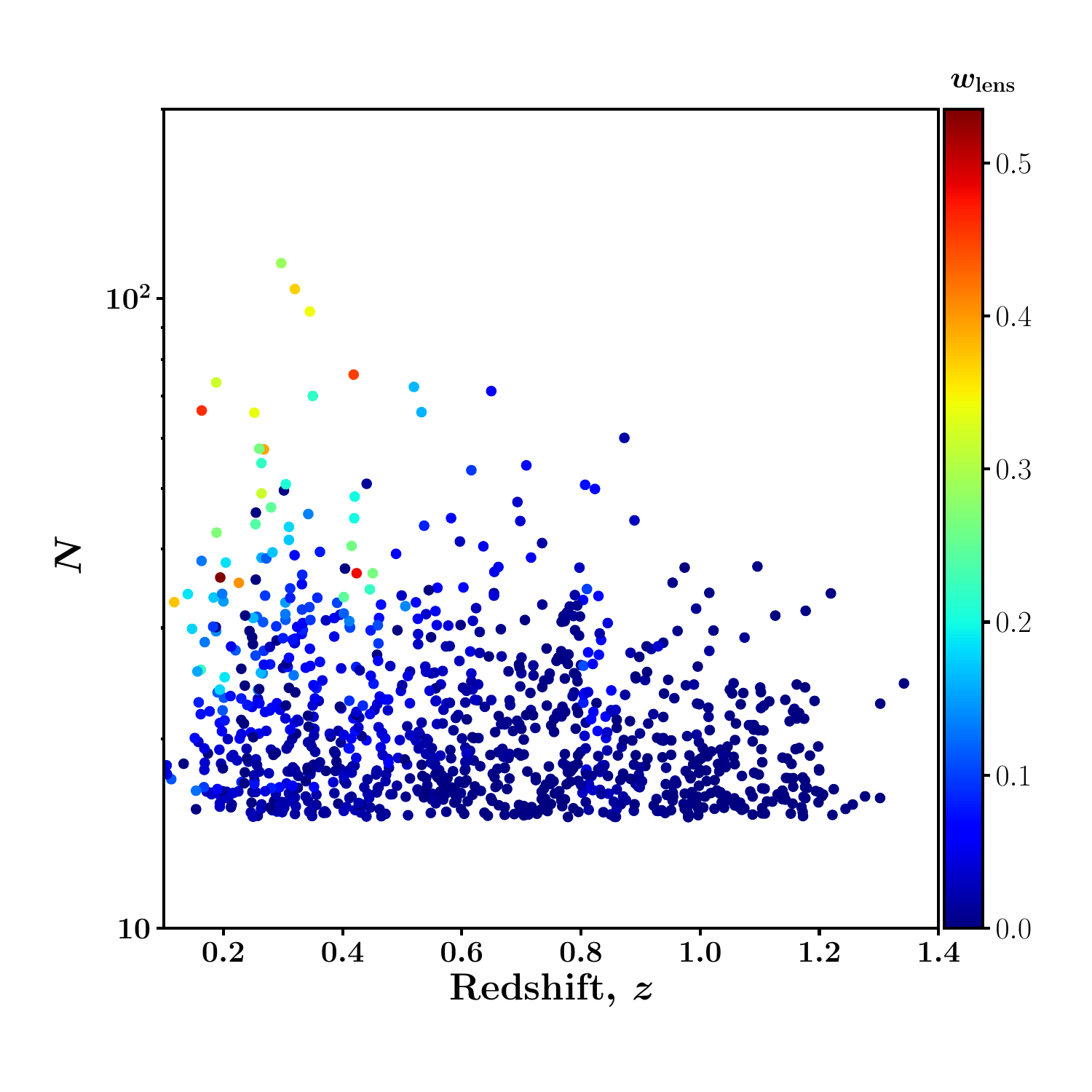}
     \caption{Lensing weight distribution in the redshift-richness plane. Each point is color-coded by the lensing weight $w_{\rm lens}$.}
   \label{fig:z_N_w}
\end{figure}

\section{Posterior distributions of the scaling relations}
\label{appendix:posterior}

Figs.~\ref{fig:posterior-distribution-1} - \ref{fig:posterior-distribution-3} display the posterior distributions of the fitted scaling relation parameters (Table \ref{tab:scaling2}, rows 1--2 and Table \ref{tab:scaling_xray_detected}). These plots illustrate the posterior median, standard deviation, and the overall parameter covariances, providing a clear view of the central tendency and dispersion of each fit.

\begin{figure*}
    \centering
    \includegraphics[width=0.9\textwidth]{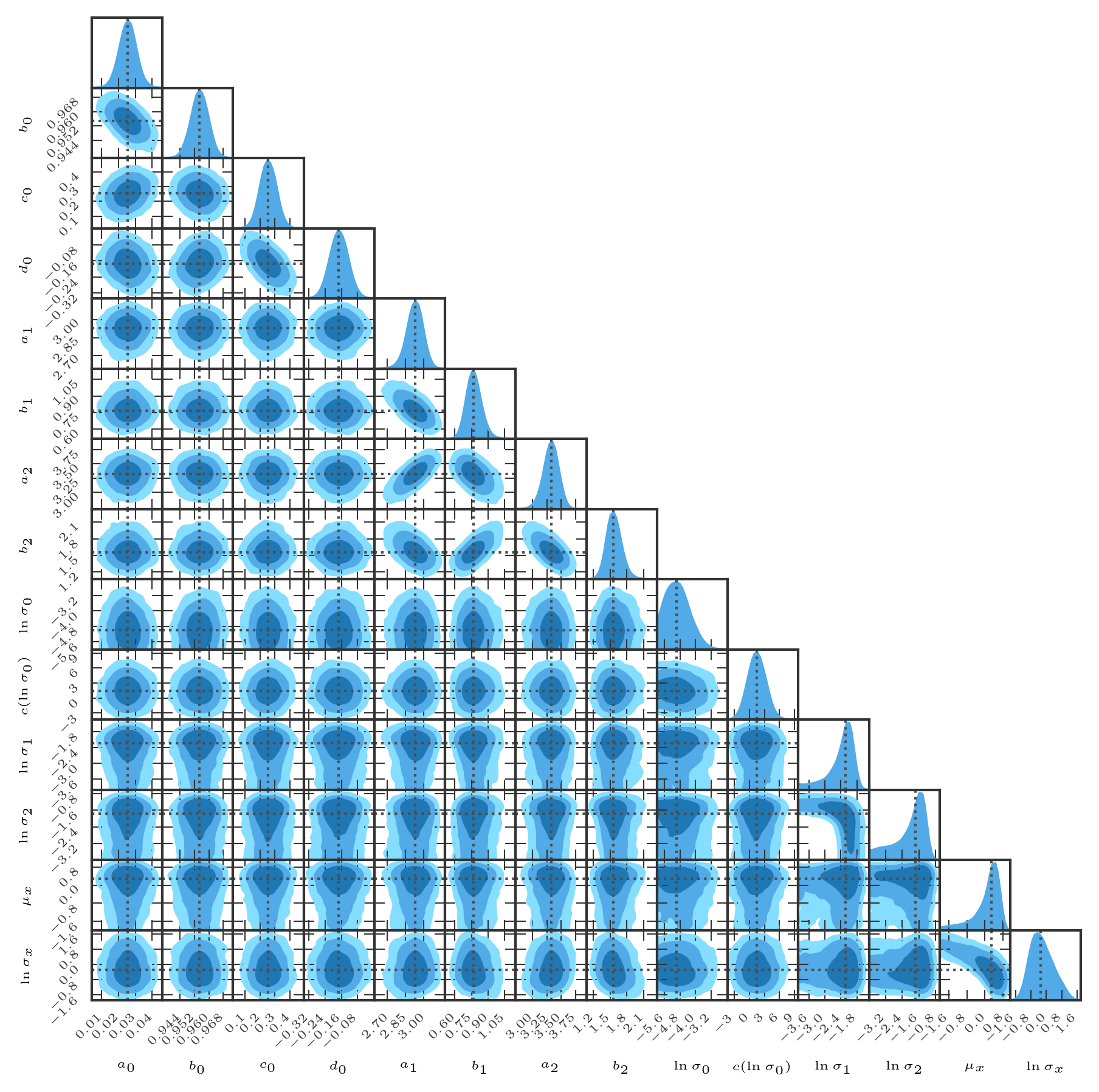}
    \caption{Posterior distributions of the scaling relation parameters (Table~\ref{tab:scaling2}, rows 1-2). Diagonal panels show the 1D marginalized posteriors, while off-diagonal panels show the joint posteriors with contours enclosing the 68\%, 95\%, 99.7\% credible regions (blue areas). Black dashed lines mark the reported best-fit values. Parameters $a$, $b$, $c$, and $d$ denote the intercepts, slopes, and redshift-evolution coefficients in Eq.~\ref{eq:scaling}. $\ln\sigma$ shows the intrinsic log-scatter of the observables, and $\mu_{x}$ and $\ln\sigma_{x}$ are the mean and log-scatter of the latent true mass distribution at the pivot redshift.
    }\label{fig:posterior-distribution-1}
\end{figure*}

\begin{figure*}
    \centering
    \includegraphics[width=0.9\textwidth]{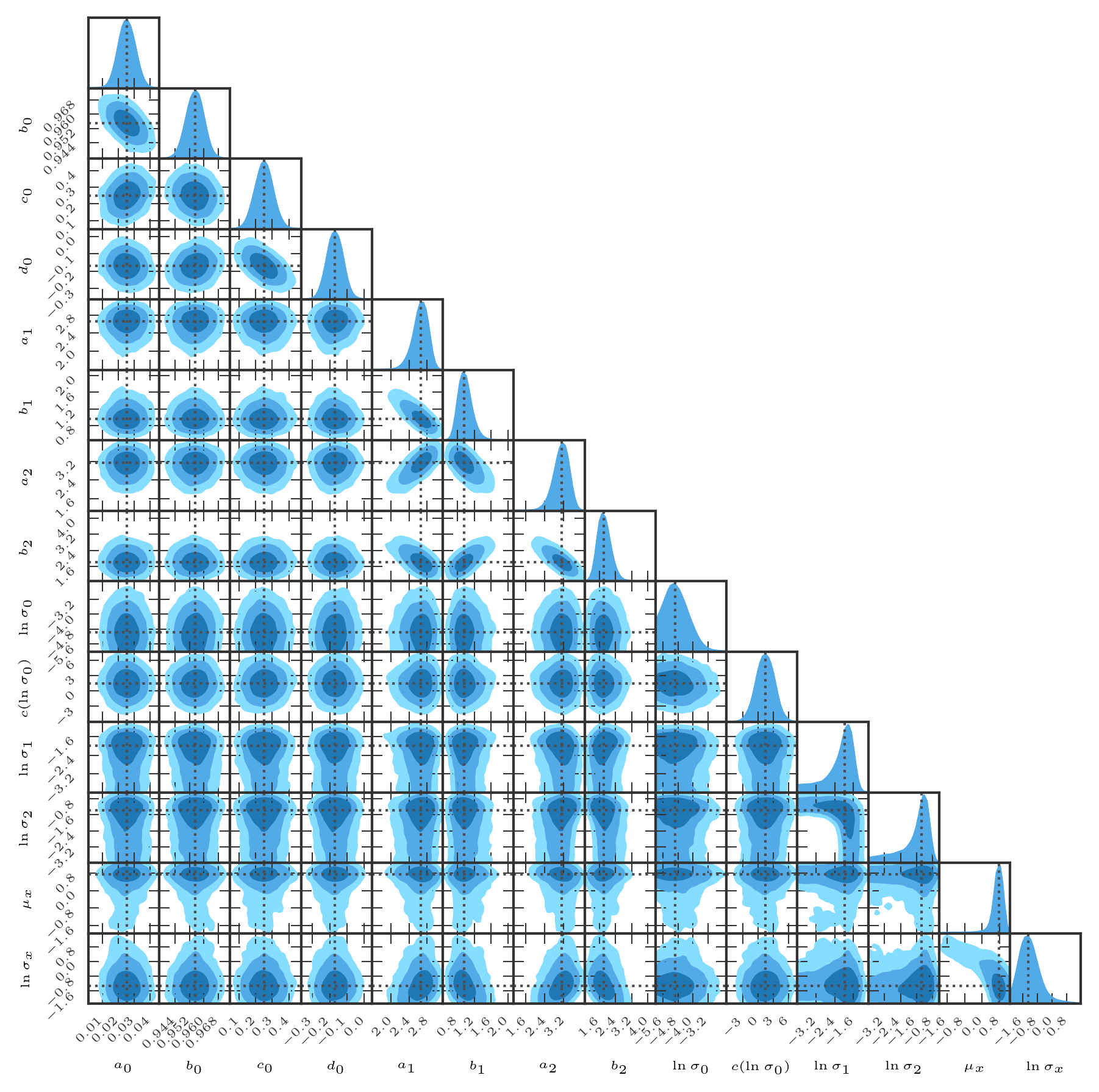}
    \caption{Same as Fig. \ref{fig:posterior-distribution-1} but for Table~\ref{tab:scaling_xray_detected}, row 1--2.
    }\label{fig:posterior-distribution-2}
\end{figure*}

\begin{figure*}
    \centering
    \includegraphics[width=0.9\textwidth]{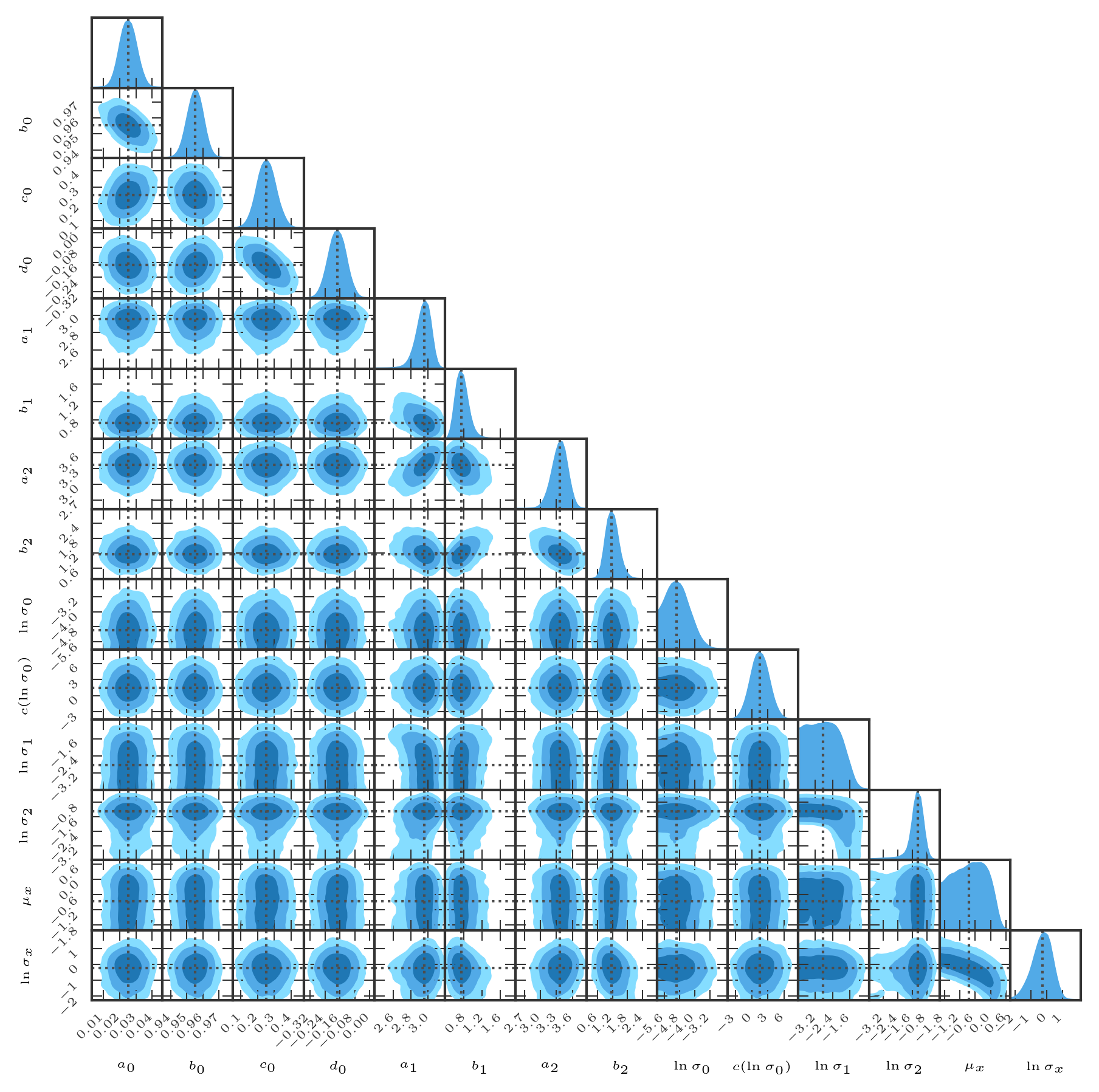}
    \caption{Same as Fig. \ref{fig:posterior-distribution-1} but for Table~\ref{tab:scaling_xray_detected}, row 3--4.
    }\label{fig:posterior-distribution-3}
\end{figure*}

\section{The richness - luminosity relation} \label{app:N-L}
For completeness, we also describe the simplified one-dimensional {\tt HiBRECS} routine, in which a single observable is regressed against mass. This version is useful for validation and for cases where only a single observable is available, but is not used in our main results in Sect.~\ref{subsec:scaling}. We apply this routine to investigate the relation between optical richness and bolometric X-ray luminosity for the stacked clusters (Sect.~\ref{subsec:stack_lumi}). The best-fit parameters are listed in Table~\ref{Table-B1}. As shown in Fig.~\ref{fig:scaling4}, the stacked richness and luminosity show a tight correlation.

At high richness ($N \gtrsim 40$), our baseline agrees well with the results of \citet{2023A&A...669A.110O}. In the lower richness regime ($N < 40$), our best-fit relation lies below that of \citeauthor{2023A&A...669A.110O}, though the two results are consistent within $1\sigma$ uncertainties.

The positive correlation between X-ray luminosity and richness indicates that richer clusters tend to be more X-ray luminous, consistent with previous findings \citep[e.g.,][]{2022MNRAS.511.2968P}.

\begin{table*}[htb]
\renewcommand{\arraystretch}{1.5}
   \caption{$N-L$ scaling relations of the stacked CAMIRA clusters.}\label{Table-B1}
    \centering
    \begin{tabular}{llllllll} \hline\hline
Relation &  $a$ & $b$ & $c$ & $d$ &$\sigma$ & $\Delta$AIC & $\Delta$BIC\\ \hline
$NE(z) - LE(z)^{-1}$ & $0.565_{-0.059}^{+0.055}$ &  $0.488_{-0.029}^{+0.030}$ & 0 & 0 & $0.133_{-0.017}^{+0.022}$ &  0 &0\\
$NE(z) - LE(z)^{-1}$ & $0.558_{-0.061}^{+0.056}$ &  $0.491_{-0.028}^{+0.031}$ & $-0.11_{-1.45}^{+1.38}$ &  $0.028_{-0.321}^{+0.325}$ &  $0.138_{-0.018}^{+0.026}$  & $+4.3$ & $+8.6$\\
\hline
\end{tabular}
%
     \tablefoot{
    Same as Table~\ref{tab:scaling2}, but for the $N-L$ scaling relations of the stacked CAMIRA clusters.
    }

\end{table*}

\begin{figure}
    \centering
    \includegraphics[width=0.48\textwidth]{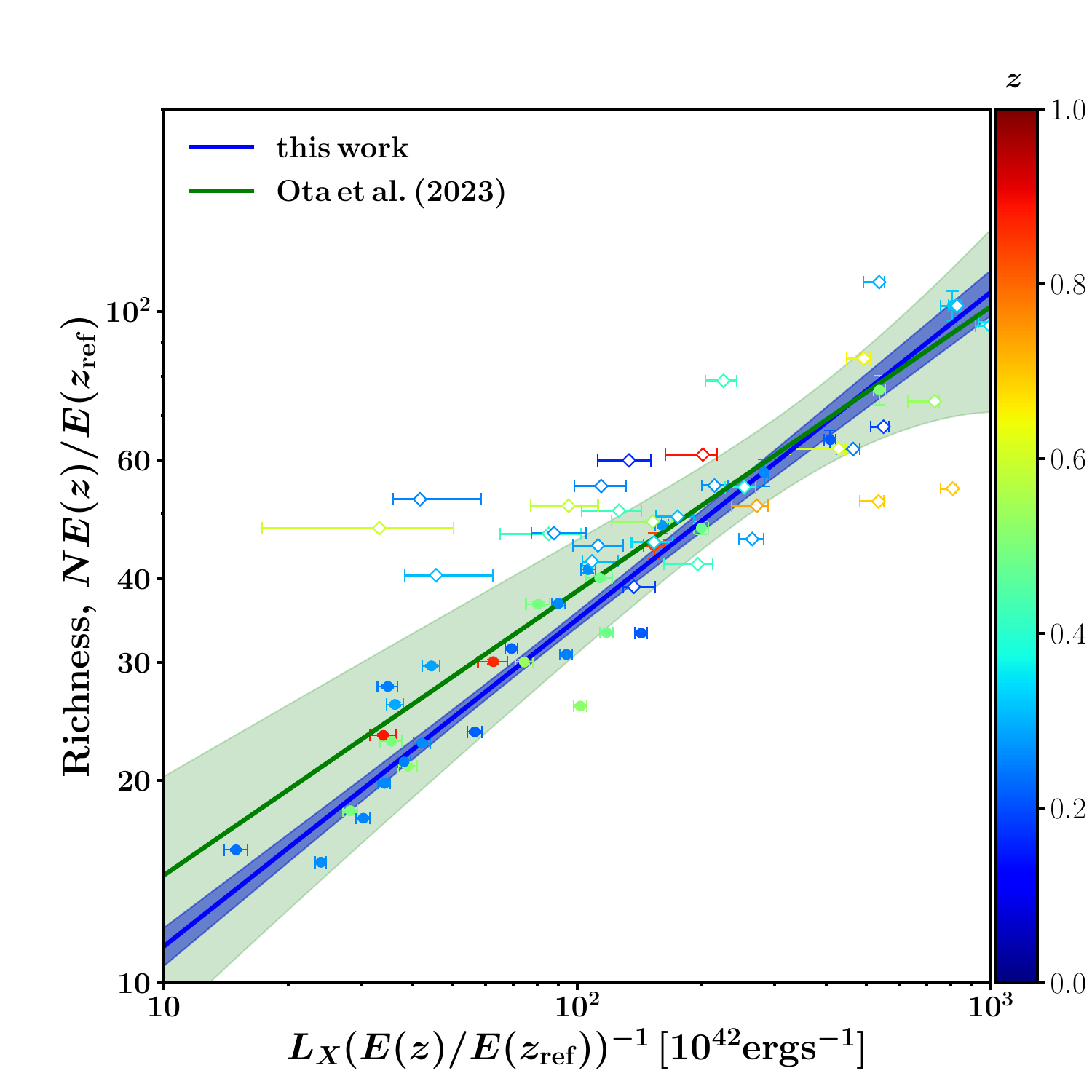}
    \caption{
    Richness versus bolometric X-ray luminosity for the stacked cluster subsamples. Colored circles indicate the stacked quantities, with the color representing the redshift of each subsample. Open diamonds show the results from \citet{2023A&A...669A.110O}. The blue and green lines indicate the best-fit scaling relations from this study and from \cite{2023A&A...669A.110O}, respectively. 
    }\label{fig:scaling4}
\end{figure}

\begin{acknowledgements}

This work is based on data from eROSITA, the soft X-ray instrument aboard SRG, a joint Russian-German science mission supported by the Russian Space Agency (Roskosmos), in the interests of the Russian Academy of Sciences represented by its Space Research Institute (IKI), and the Deutsches Zentrum f\"{u}r Luft- und Raumfahrt (DLR). The SRG spacecraft was built by Lavochkin Association (NPOL) and its subcontractors, and is operated by NPOL with support from the Max Planck Institute for Extraterrestrial Physics (MPE).

The development and construction of the eROSITA X-ray instrument was led by MPE, with contributions from the Dr. Karl Remeis Observatory Bamberg \& ECAP (FAU Erlangen-Nuernberg), the University of Hamburg Observatory, the Leibniz Institute for Astrophysics Potsdam (AIP), and the Institute for Astronomy and Astrophysics of the University of T\"{u}bingen, with the support of DLR and the Max Planck Society. The Argelander Institute for Astronomy of the University of Bonn and the Ludwig Maximilians Universit\"{a}t Munich also participated in the science preparation for eROSITA.

The eROSITA data shown here were processed using the eSASS/NRTA software system developed by the German eROSITA consortium.

The Hyper Suprime-Cam (HSC) collaboration includes the astronomical communities of Japan and Taiwan, and Princeton University. The HSC instrumentation and software were developed by the National Astronomical Observatory of Japan(NAOJ), the Kavli Institute for the Physics and Mathematics of the Universe (Kavli IPMU), the University of Tokyo, the High Energy Accelerator Research Organization (KEK), the Academia Sinica Institute for Astronomy and Astrophysics in Taiwan (ASIAA), and Princeton University. Funding was contributed by the FIRST program from Japanese Cabinet Office, the Ministry of Education, Culture, Sports, Science and Technology (MEXT), the Japan Society for the Promotion of Science (JSPS), Japan Science and Technology Agency (JST), the Toray Science Foundation, NAOJ, Kavli IPMU, KEK,ASIAA, and Princeton University.

This work was supported in part by the Fund for the Promotion of Joint International Research, JSPS KAKENHI Grant Number 16KK0101, 20K04027(NO), 20H05856, 20H00181, 19KK0076(NO and MO), 22H01260 (MO). NO acknowledges partial support by the Organization for the Promotion of Gender Equality at Nara Women's University. VG acknowledges the financial contribution from the contracts Prin-MUR 2022 supported by Next Generation EU (M4.C2.1.1, n.20227RNLY3 {\it The concordance cosmological model: stress-tests with galaxy clusters})
\end{acknowledgements}

\end{document}